\documentstyle[12pt,epsf]{article}
\newcommand{\be}{\begin{equation}}
\newcommand{\ee}{\end{equation}}

\newcommand{\beqa}{\begin{eqnarray}}
\newcommand{\eeqa}{\end{eqnarray}}
\newcommand{\nn}{\nonumber}

\newcommand{\eqref}[1]{(\ref{#1})}

\def\boxit#1{\vbox{\hrule\hbox{\vrule\kern8pt
\vbox{\hbox{\kern8pt}\hbox{\vbox{#1}}\hbox{\kern8pt}}
\kern8pt\vrule}\hrule}}
\def\mathboxit#1{\vbox{\hrule\hbox{\vrule\kern8pt\vbox{\kern8pt
\hbox{$\displaystyle #1$}\kern8pt}\kern8pt\vrule}\hrule}}

\def\IB{\relax\hbox{$\inbar\kern-.3em{\rm B}$}}
\def\IC{\relax\hbox{$\inbar\kern-.3em{\rm C}$}}
\def\ID{\relax\hbox{$\inbar\kern-.3em{\rm D}$}}
\def\IE{\relax\hbox{$\inbar\kern-.3em{\rm E}$}}
\def\IF{\relax\hbox{$\inbar\kern-.3em{\rm F}$}}
\def\IG{\relax\hbox{$\inbar\kern-.3em{\rm G}$}}
\def\IGa{\relax\hbox{${\rm I}\kern-.18em\Gamma$}}
\def\IH{\relax{\rm I\kern-.18em H}}
\def\IK{\relax{\rm I\kern-.18em K}}
\def\IL{\relax{\rm I\kern-.18em L}}
\def\IP{\relax{\rm I\kern-.18em P}}
\def\IR{\relax{\rm I\kern-.18em R}}
\def\IZ{\relax\ifmmode\mathchoice
{\hbox{\cmss Z\kern-.4em Z}}{\hbox{\cmss Z\kern-.4em Z}}
{\lower.9pt\hbox{\cmsss Z\kern-.4em Z}} {\lower1.2pt\hbox{\cmsss
Z\kern-.4em Z}}\else{\cmss Z\kern-.4em Z}\fi}

\def\II{\relax{\rm I\kern-.18em I}}

\def\CA {{\cal A}}

\def\CG {{\cal G}}

\def\CL {{\cal L}}

\begin{document}

\hfill  NRCPS-HE-05-55

\vspace{1cm}
\begin{center}
{\LARGE ~\\ {\it Non-Abelian Tensor Gauge Fields}\\
{\it and } \\
{\it Extended Current Algebra}\\
{\large ~\\Generalization of Yang-Mills Theory}

}

\vspace{1cm}

{\sl George Savvidy\\
Demokritos National Research Center\\
Institute of Nuclear Physics\\
Ag. Paraskevi, GR-15310 Athens,Greece  \\
\centerline{\footnotesize\it E-mail: savvidy@inp.demokritos.gr}
}
\end{center}
\vspace{60pt}

\centerline{{\bf Abstract}}

\vspace{12pt}

\noindent
We suggest an infinite-dimensional extension of the gauge
transformations which includes
non-Abelian tensor gauge fields. Extended gauge transformations
of non-Abelian tensor gauge fields form a new large group
which has natural geometrical interpretation it terms
of extended current algebra associated with compact Lie group.
We shall demonstrate that one can construct two infinite
series of gauge invariant quadratic forms, so that a linear
combination of them comprises the general Lagrangian.
The general Lagrangian exhibits enhanced local gauge invariants
with double number of gauge parameters and allows to eliminate
all negative norm states of the nonsymmetric second-rank tensor
gauge field.   Therefore
it describes two polarizations of helicity-two and
helicity-zero massless charged tensor gauge bosons.


\newpage

\pagestyle{plain}

\section{{\it Introduction}}

The non-Abelian local gauge invariance, which was formulated
by Yang and Mills in \cite{yang},
requires that all interactions must be invariant under
independent rotations of internal
charges at all space-time points.
The gauge principle allows very little arbitrariness: the interaction of matter
fields which carry non-commuting internal charges and the nonlinear
self-interaction of gauge bosons are essentially fixed by the requirement
of local gauge invariance, very similar to the self-interaction of
gravitons in general relativity.

It is therefore appealing to extend the gauge principle, which
was elevated by Yang and
Mills to a powerful constructive principle, so that it will define the
interaction of matter fields which carry
not only non-commutative internal charges, but
also arbitrary large spins. It seems that this will naturally
lead to a theory in which fundamental forces will be mediated by
integer-spin gauge quanta  and
that the Yang-Mills vector gauge boson will become a member of a bigger
family of tensor gauge bosons.

In the previous papers \cite{Savvidy:2005fi,Savvidy:2005zm} we extended the
gauge principle so that it enlarges
the original algebra of the Abelian local gauge transformations
found in \cite{Savvidy:dv,Savvidy:2003fx,Savvidy:2005fe}
to a non-Abelian case. The extended non-Abelian gauge
transformations of the tensor gauge fields form {\it a new large group
which has a natural geometrical interpretation in terms of extended current
algebra associated with the Lorentz group}.
On this large group one can define field strength tensors,
which are transforming homogeneously with respect to the extended
gauge transformations.  The invariant Lagrangian is quadratic in
the field strength tensors and describes interaction of tensor gauge
fields of arbitrary large integer spin $1,2,...$.

We shall present a second invariant Lagrangian
which can be constructed in terms of the above field strength tensors.
The total Lagrangian is a sum of the two Lagrangians and
exhibits enhanced local gauge invariance with double number of gauge parameters.
This allows to eliminate all negative norm states of the nonsymmetric second rank
tensor gauge field $A_{\mu\lambda}$, which describes therefore
two polarizations of helicity-two massless charged tensor gauge boson and
of the helicity-zero "axion".

The early investigation of higher-spin representations of the Poincar\'e
algebra and of the corresponding field equations is due to Majorana,
Dirac and Wigner \cite{majorana,dirac,wigner}. The theory of massive particles
of higher spin was further developed by Fierz and Pauli \cite{fierzpauli} and
Rarita and Schwinger \cite{rarita}. The Lagrangian and
S-matrix formulations of {\it free field
theory} of massive and massless fields with higher spin
have been completely constructed in
\cite{yukawa1,schwinger,Weinberg:1964cn,Weinberg:1964ev,Weinberg:1964ew,
chang,singh,singh1,fronsdal,fronsdal1}.
The problem of {\it introducing interaction} appears to be much more complex
\cite{Gupta,kraichnan,thirring,feynman,deser,fronsdal2,Sezgin:2001zs,Sagnotti:2005ns}
and met enormous difficulties for spin fields higher than two
\cite{witten,deser1,berends,dewit,vasiliev}.
The first positive result in this direction was
the light-front construction of the cubic
interaction term for the massless field of helicity $\pm \lambda$ in
\cite{Bengtsson:1983pd,Bengtsson:1983pg}.

In our approach the gauge fields are defined as rank-$(s+1)$ tensors
\cite{Savvidy:2005fi,Savvidy:2005zm}
$$
A^{a}_{\mu\lambda_1 ... \lambda_{s}}(x),~~~~~s=0,1,2,...
$$
and are totally symmetric with respect to the
indices $  \lambda_1 ... \lambda_{s}  $. {\it A priory} the tensor gauge fields
have no symmetries with
respect to the first index  $\mu$. The index $a$ numerates the generators $L^a$
of the Lie algebra $\breve{g}$ of a {\it compact}\footnote{The algebra $\breve{g}$
possesses an orthogonal
basis in which the structure constant $f^{abc}$ are totally antisymmetric.}
Lie group G.
One can think of these tensor gauge fields
as appearing in the expansion of the extended gauge field $\CA_{\mu}(x,e)$
over the coordinates $e^{\lambda}$  of the tangent space
\cite{Savvidy:2005fi,Savvidy:2005zm,Savvidy:2003fx}:
\be\label{gaugefield}
\CA_{\mu}(x,e) = \sum^{\infty}_{s=0}~{1 \over s!}
A^{a}_{\mu\lambda_1 ... \lambda_{s}}(x) ~L^a e^{\lambda_1}...e^{\lambda_s}.
\ee
In this sense the gauge field $A^{a}_{\mu\lambda_1 ... \lambda_{s}}$  carries extra
indices $\lambda_1, ..., \lambda_{s}$, which together with index $a$
are labeling the generators $L^{a}_{\lambda_1 ... \lambda_{s}}$  of
{\it extended current
algebra $\CG$ associated with the Lorentz group.} The corresponding algebra
has infinite many generators
$L^{a}_{\lambda_1 ... \lambda_{s}} = L^a e_{\lambda_1}...e_{\lambda_s}$ and
 is given by the commutator
\be
[L^{a}_{\lambda_1 ... \lambda_{s}}, L^{b}_{\rho_1 ... \rho_{k}}]=if^{abc}
L^{c}_{\lambda_1 ... \lambda_{s}\rho_1 ... \rho_{k}}.
\ee

{\it The extended non-Abelian gauge transformations of the tensor gauge
fields are defined
by the following equations } \cite{Savvidy:2005fi,Savvidy:2005zm}:
\beqa\label{polygauge}
\delta A^{a}_{\mu} &=& ( \delta^{ab}\partial_{\mu}
+g f^{acb}A^{c}_{\mu})\xi^b ,~~~~~\\
(I)~~~~~~~\delta A^{a}_{\mu\nu} &=&  ( \delta^{ab}\partial_{\mu}
+  g f^{acb}A^{c}_{\mu})\xi^{b}_{\nu} + g f^{acb}A^{c}_{\mu\nu}\xi^{b},\nonumber\\
\delta A^{a}_{\mu\nu \lambda}& =&  ( \delta^{ab}\partial_{\mu}
+g f^{acb} A^{c}_{\mu})\xi^{b}_{\nu\lambda} +
g f^{acb}(  A^{c}_{\mu  \nu}\xi^{b}_{\lambda } +
A^{c}_{\mu \lambda }\xi^{b}_{ \nu}+
A^{c}_{\mu\nu\lambda}\xi^{b}),\nn\\
.........&.&............................\nn
\eeqa
or in a general form by the formula
\be\label{generalgaugetransform}
\delta A^{a}_{\mu\lambda_1 ... \lambda_s} = ( \delta^{ab}\partial_{\mu}
+g f^{acb} A^{c}_{\mu})\xi^{b}_{\lambda_1\lambda_2 ...\lambda_s} +
g f^{acb}~\sum^{s}_{i=1}  \sum_{P's} A^{c}_{\mu\lambda_1 ...\lambda_i}
\xi^{b}_{\lambda_{i+1} ...\lambda_s },
\ee
where the infinitesimal gauge parameters $\xi^{b}_{\lambda_{1} ...\lambda_s }$ are
totally symmetric rank-s tensors. The summation  $\sum_{P's}$ is
over all permutations of two
sets of indices $\lambda_1 ... \lambda_i$ and $\lambda_{i+1} ... \lambda_{s}$
which correspond to nonequal terms. It is obvious that this transformation
preserves the symmetry properties of the tensor gauge field
$A^{a}_{\mu\lambda_1 ... \lambda_s}$. Indeed, the first term in the r.h.s.
is a covariant
derivative of the totally symmetric rank-s tensor
$\nabla^{ab}_{\mu}\xi^{b}_{\lambda_1\lambda_2 ...\lambda_s}$ and every term
$\sum_{P's} A^{c}_{\mu\lambda_1 ...\lambda_i}
\xi^{b}_{\lambda_{i+1} ...\lambda_s }$ in the second sum is totally
symmetric with respect to the indices $\lambda_1\lambda_2 ...\lambda_s$ by
construction.
The matrix form of the transformation is
\beqa\label{matrixformofgaugetransformation}
\delta_{\xi} A_{\mu\lambda_1 ... \lambda_s} &=& \partial_{\mu}
\xi_{\lambda_1\lambda_2 ...\lambda_s}
-i g  [A_{\mu}, \xi_{\lambda_1\lambda_2 ...\lambda_s}] -i g
\sum^{s}_{i=1}  \sum_{P's} [A_{\mu\lambda_1 ...\lambda_i},
\xi_{\lambda_{i+1} ...\lambda_s }],
\eeqa
where the tensor gauge fields are
$A^{ab}_{\mu\lambda_1 ... \lambda_{s}} =
(L_c)^{ab}  A^{c}_{\mu\lambda_1 ... \lambda_{s}} = i f^{acb}A^{c}_{\mu
\lambda_1 ... \lambda_{s}}$,
and  $L^a$ are the generators of the compact Lie group G in the adjoint
representation.

These extended gauge transformations
generate a closed algebraic structure. To see that, one should compute the
commutator of two extended gauge transformations $\delta_{\eta}$ and $\delta_{\xi}$
of parameters $\eta$ and $\xi$.
The commutator of two transformations can be expressed in the
form \cite{Savvidy:2005fi,Savvidy:2005zm}
\be\label{gaugecommutator}
[~\delta_{\eta},\delta_{\xi}]~A_{\mu\lambda_1\lambda_2 ...\lambda_s} ~=~
-i g~ \delta_{\zeta} A_{\mu\lambda_1\lambda_2 ...\lambda_s}
\ee
and is again an extended gauge transformation with the gauge parameters
$\{\zeta\}$ which are given by the matrix commutators
\beqa\label{gaugealgebra}
\zeta&=&[\eta,\xi]\\
\zeta_{\lambda_1}&=&[\eta,\xi_{\lambda_1}] +[\eta_{\lambda_1},\xi]\nn\\
\zeta_{\nu\lambda} &=& [\eta,\xi_{\nu\lambda}] +  [\eta_{\nu},\xi_{\lambda}]
+ [\eta_{\lambda},\xi_{\nu}]+[\eta_{\nu\lambda},\xi],\nn\\
......&.&..........................\nn
\eeqa
{\it The generalized field strengths  are defined as}
\cite{Savvidy:2005fi,Savvidy:2005zm}
\beqa\label{fieldstrengthparticular}
G^{a}_{\mu\nu} &=&
\partial_{\mu} A^{a}_{\nu} - \partial_{\nu} A^{a}_{\mu} +
g f^{abc}~A^{b}_{\mu}~A^{c}_{\nu},\\
G^{a}_{\mu\nu,\lambda} &=&
\partial_{\mu} A^{a}_{\nu\lambda} - \partial_{\nu} A^{a}_{\mu\lambda} +
g f^{abc}(~A^{b}_{\mu}~A^{c}_{\nu\lambda} + A^{b}_{\mu\lambda}~A^{c}_{\nu} ~),\nn\\
G^{a}_{\mu\nu,\lambda\rho} &=&
\partial_{\mu} A^{a}_{\nu\lambda\rho} - \partial_{\nu} A^{a}_{\mu\lambda\rho} +
g f^{abc}(~A^{b}_{\mu}~A^{c}_{\nu\lambda\rho} +
 A^{b}_{\mu\lambda}~A^{c}_{\nu\rho}+A^{b}_{\mu\rho}~A^{c}_{\nu\lambda}
 + A^{b}_{\mu\lambda\rho}~A^{c}_{\nu} ~),\nn\\
 ......&.&............................................\nn
\eeqa
and transform homogeneously with respect to the extended
gauge transformations (\ref{polygauge}). The field strength tensors are
antisymmetric in their first two indices and are totaly symmetric with respect to the
rest of the indices. By induction the entire construction
can be generalized to include tensor fields of any rank s, and
the corresponding field strength we shall define by the following expression:
\be\label{fieldstrengthgeneral}
G^{a}_{\mu\nu ,\lambda_1 ... \lambda_{s}} =
\partial_{\mu} A^{a}_{\nu \lambda_1 ... \lambda_{s}} -
\partial_{\nu} A^{a}_{\mu \lambda_1 ... \lambda_s} +
g f^{abc}\sum^{s}_{i=0}~ \sum_{P's} ~A^{b}_{\mu \lambda_1 ... \lambda_i}~
A^{c}_{\nu \lambda_{i+1} ... \lambda_{s}},
\ee
where the sum $\sum_{P's}$ runs over all permutations of two
sets of indices $\lambda_1 ... \lambda_i$ and $\lambda_{i+1} ... \lambda_{s}$
which correspond to nonequal terms.
All permutations of indices within two sets $\lambda_1 ... \lambda_i$ and
$\lambda_{i+1} ... \lambda_{s}$ correspond to equal terms, because
gauge fields are totally symmetric with respect to $\lambda_1 ... \lambda_i$ and
$\lambda_{i+1} ... \lambda_{s}$. Therefore there are
$$
\frac{s!}{i!(s-i)!}
$$
nonequal terms in the sum $\sum_{P's}$. Thus in the sum $\sum_{P's}$ there is one
term of the form $A_{\mu}A_{\nu\lambda_1\lambda_{2} ... \lambda_{s}}$, there are
$s$ terms
of the form $A_{\mu \lambda_1}
A_{\nu \lambda_{2} ... \lambda_{s}}$ and $s(s-1)/2$ terms of the form
$A_{\mu \lambda_1 \lambda_2}~
A_{\nu \lambda_{3} ... \lambda_{s}}$ and so on.
In the above definition of the extended gauge field strength
$G^{a}_{\mu\nu ,\lambda_1 ... \lambda_{s}}$,
together with the classical Yang-Mills gauge boson
$A^{a}_{\mu}$, there participate a set of higher-rank gauge fields
$A^{a}_{\mu\lambda_1}$, $A^{a}_{\mu\lambda_1 , \lambda_2}$,  ... ,
$A^{a}_{\mu\lambda_1 ...\lambda_{s}}$ up to the rank $s+1$.
By construction the field strength (\ref{fieldstrengthgeneral})
is antisymmetric with respect to its first two indices
$
G^{a}_{\mu\nu ,\lambda_1 ... \lambda_{s}}~ = ~-~G^{a}_{\nu \mu,\lambda_1 ... \lambda_{s}}
$
and is totally symmetric with respect to the rest of the indices
$
G^{a}_{\mu\nu ,..\lambda_{i}...\lambda_{j}.. } =
G^{a}_{\mu\nu ,..\lambda_{j}...\lambda_{i}..}~,
$
where $1 \leq i < j \leq s$.

The inhomogeneous extended gauge transformation (\ref{generalgaugetransform})
induces the homogeneous gauge
transformation of the corresponding field strength
(\ref{fieldstrengthgeneral}) of the form \cite{Savvidy:2005fi,Savvidy:2005zm}
\beqa\label{fieldstrenghparticular}
\delta G^{a}_{\mu\nu}&=& g f^{abc} G^{b}_{\mu\nu} \xi^c \\
\delta G^{a}_{\mu\nu,\lambda} &=& g f^{abc} (~G^{b}_{\mu\nu,\lambda} \xi^c
+ G^{b}_{\mu\nu} \xi^{c}_{\lambda}~),\nonumber\\
\delta G^{a}_{\mu\nu,\lambda\rho} &=& g f^{abc}
(~G^{b}_{\mu\nu,\lambda\rho} \xi^c
+ G^{b}_{\mu\nu,\lambda} \xi^{c}_{\rho} +
G^{b}_{\mu\nu,\rho} \xi^{c}_{\lambda} +
G^{b}_{\mu\nu} \xi^{c}_{\lambda\rho}~)\nn\\
......&.&..........................,\nn
\eeqa
or in general
\be\label{variationfieldstrengthgeneral}
\delta G^{a}_{\mu\nu,\lambda_1 ... \lambda_s} =
g f^{abc} \sum^{s}_{i=0}~ \sum_{P's} ~G^{b}_{\mu\nu,\lambda_1 ... \lambda_i}
\xi^{c}_{\lambda_{i+1}...\lambda_s}.
\ee
The symmetry properties of the field strength
$G^{a}_{\mu\nu,\lambda_1 ... \lambda_s}$
remain invariant in the course of this transformation.

The gauge invariant Lagrangian  now can be formulated in the
form \cite{Savvidy:2005fi,Savvidy:2005zm}
\beqa\label{fulllagrangian1}
{{\cal L}}_{s+1}&=&-{1\over 4} ~
G^{a}_{\mu\nu, \lambda_1 ... \lambda_s}~
G^{a}_{\mu\nu, \lambda_{1}...\lambda_{s}} +.......\nonumber\\
&=& -{1\over 4}\sum^{2s}_{i=0}~a^{s}_i ~
G^{a}_{\mu\nu, \lambda_1 ... \lambda_i}~
G^{a}_{\mu\nu, \lambda_{i+1}...\lambda_{2s}}
(\sum_{p's} \eta^{\lambda_{i_1} \lambda_{i_2}} .......
\eta^{\lambda_{i_{2s-1}} \lambda_{i_{2s}}})~,
\eeqa
where the sum $\sum_p$ runs over all nonequal permutations of $i's$,
in total $(2s-1)!!$
terms. For the low
values of $s=0,1,2,...$ the numerical coefficients and eta functions are
$$
a^{s}_i = {s!\over i!(2s-i)!}
$$
$a^{0}_0=1;~~a^{1}_1 =1,a^{1}_0 =a^{1}_2 =1/2;~~
a^{2}_2 =1/2,a^{2}_1 =a^{2}_3 =1/3,a^{2}_0 =a^{2}_4 =1/12$ and
$\eta^{\lambda_{1} \lambda_{2} },~ \eta^{\lambda_{1} \lambda_{2}}
\eta^{\lambda_{3} \lambda_{4}}+\eta^{\lambda_{1} \lambda_{3}}
\eta^{\lambda_{2} \lambda_{4}}+\eta^{\lambda_{1} \lambda_{4}}
\eta^{\lambda_{2} \lambda_{3}}, ...$ and so on.
In order to describe fixed rank-$(s+1)$ gauge field
one should have  at disposal all gauge fields
up to the rank $2s+1$.
In order to make all tensor gauge fields dynamical one should add
the corresponding kinetic terms. Thus the invariant
Lagrangian describing dynamical tensor gauge bosons of all ranks
has the form
\be\label{fulllagrangian2}
{{\cal L}} = \sum^{\infty}_{s=1}~ g_s {{\cal L}}_s~.
\ee
The first three terms of the invariant Lagrangian have
the following form \cite{Savvidy:2005fi,Savvidy:2005zm}:
\beqa\label{firstthreeterms}
{{\cal L}} =  {{\cal L}}_1 +  {{\cal L}}_2 +{{\cal L}}_3 +...
=-{1\over 4}G^{a}_{\mu\nu}
G^{a}_{\mu\nu}
-{1\over 4}G^{a}_{\mu\nu,\lambda}G^{a}_{\mu\nu,\lambda}
-{1\over 4}G^{a}_{\mu\nu}G^{a}_{\mu\nu,\lambda\lambda}~-~~~~~~~~~~\nn\\
-{1\over 4}G^{a}_{\mu\nu,\lambda\rho}G^{a}_{\mu\nu,\lambda\rho}
-{1\over 8}G^{a}_{\mu\nu ,\lambda\lambda}G^{a}_{\mu\nu ,\rho\rho}
-{1\over 2}G^{a}_{\mu\nu,\lambda}  G^{a}_{\mu\nu ,\lambda \rho\rho}
-{1\over 8}G^{a}_{\mu\nu}  G^{a}_{\mu\nu ,\lambda \lambda\rho\rho}~+..,
\eeqa
where the first term is the Yang-Mills Lagrangian and the
second and the third ones describe the tensor gauge fields $A^{a}_{\mu\nu},
A^{a}_{\mu\nu\lambda}$ and so on.
It is important that:  i) {\it the Lagrangian does not
contain higher derivatives of tensor gauge fields
ii) all interactions take place
through the three- and four-particle exchanges with dimensionless
coupling constant  iii) the complete Lagrangian contains all higher-rank
tensor gauge fields and should not be truncated}.

\section{{\it Geometrical Interpretation}}

Let us consider a possible geometrical interpretation of the
above construction. Introducing the coordinates $e^{\mu}$ on the tangent
space we shall consider non-Abelian gauge transformations $U(\xi)$ with
the extended gauge parameter  $\xi(x,e) $ which
depends on the space-time coordinates $x^{\mu}$ and the tangent
coordinates $e^{\lambda}$.  We can expand the gauge parameter $\xi(x,e) $
in series using generators
$L^{a}_{\lambda_1...\lambda_s} = L^a e_{\lambda_1}...e_{\lambda_s}$
\cite{Savvidy:2005fi,Savvidy:2005zm,Savvidy:2003fx}
\be
\xi(x,e)=   \sum^{\infty}_{s=0}~{1\over s!}
\xi^{a}_{\lambda_1 ... \lambda_{s}}(x) ~L^a   e_{\lambda_1}...e_{\lambda_s}
\ee
and define the gauge transformation of the extended gauge field
$\CA_{\mu}(x,e)$   as in (\ref{polygauge})
\be\label{extendedgaugetransformation}
\CA^{'}_{\mu}(x,e) = U(\xi)  \CA_{\mu}(x,e) U^{-1}(\xi) -{i\over g}
\partial_{\mu}U(\xi) ~U^{-1}(\xi),
\ee
where the unitary transformation matrix
is given by the expression
$$
U(\xi) = exp\{ig \xi(x,e) \}.
$$
This allows to construct the extended field strength tensor of the form
(\ref{fieldstrengthparticular})
\be
\CG_{\mu\nu}(x,e) = \partial_{\mu} \CA_{\nu}(x,e) - \partial_{\nu} \CA_{\mu}(x,e) -
i g [ \CA_{\mu}(x,e)~\CA_{\nu}(x,e)]
\ee
using the commutator of the covariant derivatives
$$
\nabla^{ab}_{\mu} = (\partial_{\mu}-ig \CA_{\mu}(x,e))^{ab}
$$
of a standard form
$
[\nabla_{\mu}, \nabla_{\nu}]^{ab} = g f^{acb} \CG^{c}_{\mu\nu}~,
$
so that
\be\label{fieldstrenghthtransformation}
\CG^{'}_{\mu\nu}(x,e)) = U(\xi)  \CG_{\mu\nu}(x,e) U^{-1}(\xi).
\ee
The  invariant Lagrangian density is given by the expression
\be\label{lagrangdensity}
{{\cal L}}(x,e) =  -{1\over 4} \CG^{a}_{\mu\nu}(x,e)\CG^{a}_{\mu\nu}(x,e),
\ee
as one can get convinced computing its variation with respect to the
extended gauge transformation (\ref{polygauge}),(\ref{extendedgaugetransformation})
and (\ref{fieldstrenghparticular}),(\ref{fieldstrenghthtransformation})
$$
\delta {{\cal L}}(x,e) = -{1\over 2}\CG^{a}_{\mu\nu}(x,e)~ g f^{acb}~
\CG^{c}_{\mu\nu}(x,e) ~\xi^{b}(x,e) =0.
$$
The Lagrangian density (\ref{lagrangdensity}) allows to extract
{\it gauge invariant, totally symmetric, tensor densities
$\CL_{\lambda_1 ... \lambda_{s}}(x)$}
using expansion with respect to the vector variable $e^{\lambda}$
\be
\CL(x,e) = \sum^{\infty}_{s=0}~{1\over s!}
\CL_{\lambda_1 ... \lambda_{s}}(x) ~ e_{\lambda_1}...e_{\lambda_s} .
\ee
In particular the expansion term which is quadratic in powers of $e$ is
\be
(\CL_2 )_{\lambda\rho} = -{1\over 4}G^{a}_{\mu\nu,\lambda}G^{a}_{\mu\nu,\rho}
-{1\over 4}G^{a}_{\mu\nu}G^{a}_{\mu\nu,\lambda\rho}
\ee
and defines a unique gauge invariant Lagrangian which can be
constructed from the above
tensor (see the next section for it explicit variation (\ref{explicitvariation})),
that is the Lagrangian $\CL_2$
$$
{{\cal L}}_2 =-{1\over 4}G^{a}_{\mu\nu,\lambda}G^{a}_{\mu\nu,\lambda}
-{1\over 4}G^{a}_{\mu\nu}G^{a}_{\mu\nu,\lambda\lambda}
$$
and so on.

The whole construction can be viewed as an extended vector bundle X
on which the gauge field $\CA^{a}_{\mu}(x,e)$ is a connection.
In this sense the gauge field $A^{a}_{\mu\lambda_1 ... \lambda_{s}}$  carries extra
indices $\lambda_1, ..., \lambda_{s}$, which together with index $a$
are labeling the generators $L^{a}_{\lambda_1 ... \lambda_{s}}$  of
{\it extended current
algebra $\CG$ associated with the Lorentz group.} The corresponding algebra
has infinite many generators
$L^{a}_{\lambda_1 ... \lambda_{s}} = L^a e_{\lambda_1}...e_{\lambda_s}$ and
 is given by the commutator
\be
[L^{a}_{\lambda_1 ... \lambda_{s}}, L^{b}_{\rho_1 ... \rho_{k}}]=if^{abc}
L^{c}_{\lambda_1 ... \lambda_{s}\rho_1 ... \rho_{k}}.
\ee
Thus we have vector bundle whose structure group
is an extended gauge group $\CG$ with group elements $U(\xi)=exp(i \xi(e) )$,
where $\xi(e)=  \sum_s ~
\xi^{a}_{\lambda_1 ... \lambda_{s}} ~L^a  e_{\lambda_1}...e_{\lambda_s}$
and the composition law (\ref{gaugealgebra}).
In contrast, in Kac-Moody current algebra
the generators depend on the complex variable $L^{a}_n = L^a z^n$ (see also
\cite{minnes})
$$
[L^{a}_n ,L^{b}_m] = if^{abc} L^{c}_{n+m}.
$$

In the next section we shall see that, there exist a second invariant Lagrangian
${{\cal L}}^{'}$ which can be constructed in terms of extended
field strength tensors (\ref{fieldstrengthparticular}) and the total Lagrangian is a
linear sum of the two Lagrangians $ c ~{{\cal L}} + c^{'}~ {{\cal L}}^{'} $.

\section{{\it Enhanced Local Gauge Symmetry}}

Indeed the Lagrangian (\ref{fulllagrangian1}), (\ref{fulllagrangian2})
and (\ref{firstthreeterms}) is not the
most general Lagrangian which can be constructed in terms
of the above field strength tensors
(\ref{fieldstrengthparticular}) and (\ref{fieldstrengthgeneral}).
As we shall see there exists a second invariant Lagrangian ${{\cal L}}^{'}$
(\ref{secondseries}),
(\ref{secondseriesdensities}) and (\ref{secondfulllagrangian})
which can be constructed in terms of extended
field strength tensors (\ref{fieldstrengthparticular}) and the total Lagrangian is a
linear sum of the two Lagrangians $ c ~{{\cal L}} + c^{'}~ {{\cal L}}^{'} $.
In particular for the second-rank
tensor gauge field $A^{a}_{\mu\lambda}$ the total Lagrangian is a
sum of two Lagrangians ${{\cal L}}_{2}+{{\cal L}}^{'}_{2} $ and,
with specially chosen coefficients $\{c,c^{'}\}$, it exhibits an enhanced
gauge invariance (\ref{largegaugetransformation}),(\ref{freedoublepolygaugesymmetric})
with double number of gauge parameters, which allows to eliminate negative norm
polarizations of the nonsymmetric second-rank tensor gauge field $A_{\mu\lambda}$.
The geometrical interpretation
of the enhanced gauge symmetry with double number of gauge parameters is not yet
known.

Let us consider the gauge invariant tensor density of the form
\be\label{secondseries}
{{\cal L}}^{'}_{\rho_1\rho_2}(x,e) =  {1\over 4}
\CG^{a}_{\mu\rho_1}(x,e)\CG^{a}_{\mu\rho_1}(x,e).
\ee
It is gauge invariant because its variation is also equal to zero
$$
\delta {{\cal L}}^{'}_{\rho_1\rho_2}(x,e) ={1\over 4}g f^{acb}~
\CG^{c}_{\mu\rho_1}(x,e) ~\xi^{b}(x,e)\CG^{a}_{\mu\rho_2}(x,e)+
{1\over 4}\CG^{a}_{\mu\rho_1}(x,e)~ g f^{acb}~
\CG^{c}_{\mu\rho_2}(x,e) ~\xi^{b}(x,e) =0.
$$
The Lagrangian density (\ref{secondseries}) generate the
second series of {\it gauge invariant tensor densities
$(\CL^{'}_{\rho_1\rho_2})_{\lambda_1 ... \lambda_{s}}(x)$}
when we expand it in powers of the vector variable $e$
\be\label{secondseriesdensities}
{{\cal L}}^{'}_{\rho_1\rho_2}(x,e) = \sum^{\infty}_{s=0}~{1\over s!}
(\CL^{'}_{\rho_1\rho_2})_{\lambda_1 ... \lambda_{s}}(x) ~
e_{\lambda_1}...e_{\lambda_s} .
\ee
Using contraction of these tensor densities the gauge invariant Lagrangians can
be formulated in the form
\beqa\label{secondfulllagrangian}
{{\cal L}}^{'}_{s+1}&=&{1\over 4} ~
G^{a}_{\mu\lambda_1,\lambda_2  ... \lambda_{s+1}}~
G^{a}_{\mu\lambda_2,\lambda_{1} ...\lambda_{s+1}} +{1\over 4} ~
G^{a}_{\mu\nu,\nu\lambda_3  ... \lambda_{s+1}}~
G^{a}_{\mu\rho,\rho\lambda_{3} ...\lambda_{s+1}}  +.......\nonumber\\
&=& {1\over 4}\sum^{2s+1}_{i=1}~{a^{s}_{i-1} \over s} ~
G^{a}_{\mu\lambda_1,\lambda_2  ... \lambda_i}~
G^{a}_{\mu\lambda_{i+1},\lambda_{i+2} ...\lambda_{2s+2}}
(\sum^{'}_{p's} \eta^{\lambda_{i_1} \lambda_{i_2}} .......
\eta^{\lambda_{i_{2s+1}} \lambda_{i_{2s+2}}})~,
\eeqa
where the sum $\sum^{'}_p$ runs over all nonequal permutations of $i's$,
with exclusion
of the terms which contain $\eta^{\lambda_{1},\lambda_{i+1}}$.

It is important to consider these Lagrangians in details.
The invariance of the Lagrangian
$$
{{\cal L}}_2 =-{1\over 4}G^{a}_{\mu\nu,\lambda}G^{a}_{\mu\nu,\lambda}
-{1\over 4}G^{a}_{\mu\nu}G^{a}_{\mu\nu,\lambda\lambda}
$$
in (\ref{fulllagrangian1}), (\ref{fulllagrangian2}) and (\ref{firstthreeterms})
was demonstrated in \cite{Savvidy:2005zm} by
calculating its variation with respect to the gauge
transformation (\ref{polygauge}) and (\ref{fieldstrenghparticular}),
(\ref{variationfieldstrengthgeneral}).
Indeed, its variation is equal to zero
\beqa\label{explicitvariation}
\delta \CL_2   =
&-&{1\over 4} G^{a}_{\mu\nu,\lambda} g f^{abc} (G^{b}_{\mu\nu,\lambda} \xi^c  +
G^{b}_{\mu\nu} \xi^{c}_{\lambda})
-{1\over 4}  g f^{abc} (G^{a}_{\mu\nu,\lambda}\xi^c +
G^{b}_{\mu\nu} \xi^{c}_{\lambda}   )G^{b}_{\mu\nu,\lambda}   \nn\\
&-&{1\over 4} g f^{abc} G^{b}_{\mu\nu} \xi^c G^{a}_{\mu\nu,\lambda\lambda}\nonumber\\
&-&{1\over 4} G^{a}_{\mu\nu} g f^{abc} (G^{b}_{\mu\nu,\lambda\lambda} \xi^c  +
G^{b}_{\mu\nu, \lambda} \xi^{c}_{\lambda}+G^{b}_{\mu\nu, \lambda} \xi^{c}_{\lambda}+
G^{b}_{\mu\nu} \xi^{c}_{\lambda \lambda})=0 .
\eeqa
As we have seen the Lagrangian ${{\cal L}}$   (\ref{fulllagrangian2})
is not a unique one and that there exist a second series of invariants
${{\cal L}}^{'}$ (\ref{secondfulllagrangian}). Let us construct few of them
directly, without reference to any expansion. Indeed, there are three
Lorentz invariants in our disposal
$$
G^{a}_{\mu\nu,\lambda}G^{a}_{\mu\lambda,\nu},~~~
G^{a}_{\mu\nu,\nu}G^{a}_{\mu\lambda,\lambda},~~~
G^{a}_{\mu\nu}G^{a}_{\mu\lambda,\nu\lambda}.
$$
Calculating  the variation of each of these terms with respect to
the gauge transformation (\ref{polygauge}) and (\ref{fieldstrenghparticular})
one can get convinced that a particular linear combination
\be\label{actiontwoprime}
{{\cal L}}^{'}_2 =  {1\over 4}G^{a}_{\mu\nu,\lambda}G^{a}_{\mu\lambda,\nu}
+{1\over 4}G^{a}_{\mu\nu,\nu}G^{a}_{\mu\lambda,\lambda}
+{1\over 2} G^{a}_{\mu\nu}G^{a}_{\mu\lambda,\nu\lambda}
\ee
forms an invariant form which coincides with (\ref{secondfulllagrangian}) for
s=1. The variation of the Lagrangian ${{\cal L}}^{'}_2$
under the gauge transformation (\ref{fieldstrenghparticular}) is equal to zero:
\beqa
\delta {{\cal L}}^{'}_2 =
&+&{1\over 4} G^{a}_{\mu\nu,\lambda} g f^{abc} (G^{b}_{\mu\lambda,\nu} \xi^c  +
G^{b}_{\mu\lambda} \xi^{c}_{\nu})
+{1\over 4} g f^{abc} (G^{b}_{\mu\nu,\lambda} \xi^c  +
G^{b}_{\mu\nu} \xi^{c}_{\lambda}) G^{a}_{\mu\lambda,\nu}    \nn\\
&+& {1\over 2}G^{a}_{\mu\nu,\nu}g f^{abc} (G^{b}_{\mu\lambda,\lambda}  \xi^c  +
G^{b}_{\mu\lambda} \xi^{c}_{\lambda})\nn\\
&+& {1\over 2}g f^{abc} G^{b}_{\mu\nu} \xi^c G^{a}_{\mu\lambda,\nu\lambda}\nonumber\\
&+& {1\over 2} G^{a}_{\mu\nu} g f^{abc} (G^{b}_{\mu\lambda,\nu\lambda} \xi^c  +
G^{b}_{\mu\lambda,\nu } \xi^{c}_{\lambda}+
G^{b}_{\mu\lambda,\lambda } \xi^{c}_{\nu}+
G^{b}_{\mu\lambda} \xi^{c}_{\nu \lambda})=0 .\nonumber
\eeqa
As a result we have two invariant Lagrangians
${{\cal L}}_2$ and ${{\cal L}}^{'}_2$ and the general Lagrangian is a
linear combination of these two Lagrangians
$
{{\cal L}}_2 + c {{\cal L}}^{'}_2 ,
$
where c is an arbitrary constant.

{\it Our aim now is to demonstrate  that
if $c=1$ then we shall have enhanced local gauge invariance
(\ref{largegaugetransformation}),(\ref{freedoublepolygaugesymmetric}) of
the Lagrangian
${{\cal L}}_2 + {{\cal L}}^{'}_2$ with double number of gauge parameters.
This allows to eliminate all negative norm states of the nonsymmetric second-rank
tensor gauge field $A^{a}_{\mu \lambda}$, which describes therefore two polarizations
of helicity-two  and
helicity-zero massless charged tensor gauge bosons.}

Indeed, let us consider the
situation at the linearized level when the gauge coupling constant g is equal to zero.
The free part of the ${{\cal L}}_2$ Lagrangian is
$$
{{\cal L}}^{free}_2 ={1 \over 2} A^{a}_{\alpha\acute{\alpha}}
(\eta_{\alpha\gamma}\eta_{\acute{\alpha}\acute{\gamma}}\partial^{2} -
\eta_{\acute{\alpha}\acute{\gamma}} \partial_{\alpha} \partial_{\gamma} )
A^{a}_{\gamma\acute{\gamma}} =
{1 \over 2} A^{a}_{\alpha\acute{\alpha}}
H_{\alpha\acute{\alpha}\gamma\acute{\gamma}} A^{a}_{\gamma\acute{\gamma}} ,
$$
where the quadratic form in the momentum representation has the form
$$
H_{\alpha\acute{\alpha}\gamma\acute{\gamma}}(k)=
-(k^2 \eta_{\alpha\gamma} -k_{\alpha}k_{\gamma})
\eta_{\acute{\alpha}\acute{\gamma}} =
-H_{\alpha \gamma }(k) \eta_{\acute{\alpha}\acute{\gamma}},
$$
is obviously invariant with respect to the gauge
transformation $\delta A^{a}_{\mu\lambda} =\partial_{\mu} \xi^{a}_{\lambda}$,
but it is not invariant with respect to the alternative gauge transformations
$\delta A^{a}_{\mu \lambda} =\partial_{\lambda} \eta^{a}_{\mu}$. This can be
seen, for example, from the following relations in the momentum representation:
\be\label{currentdivergence}
k_{\alpha}H_{\alpha\acute{\alpha}\gamma\acute{\gamma}}(k)=0,~~~
k_{\acute{\alpha}}H_{\alpha\acute{\alpha}\gamma\acute{\gamma}}(k)=
-(k^2 \eta_{\alpha\gamma} - k_{\alpha}k_{\gamma})k_{\acute{\gamma}} \neq 0 .
\ee
Let us consider now the free part of the second Lagrangian
\beqa
{{\cal L}}^{' free}_{2} ={1 \over 4} A^{a}_{\alpha\acute{\alpha}}
(-\eta_{\alpha\acute{\gamma}}\eta_{\acute{\alpha}\gamma}\partial^{2} -
\eta_{\alpha\acute{\alpha}}\eta_{\gamma\acute{\gamma}}\partial^{2}
+\eta_{\alpha\acute{\gamma}} \partial_{\acute{\alpha}} \partial_{\gamma}
+\eta_{\acute{\alpha}\gamma} \partial_{\alpha} \partial_{\acute{\gamma}}
+\eta_{\alpha\acute{\alpha}} \partial_{\gamma} \partial_{\acute{\gamma}}+\nn\\
+\eta_{\gamma\acute{\gamma}} \partial_{\alpha} \partial_{\acute{\alpha}}
-2\eta_{\alpha\gamma} \partial_{\acute{\alpha}} \partial_{\acute{\gamma}})
A^{a}_{\gamma\acute{\gamma}}=
{1 \over 2} A^{a}_{\alpha\acute{\alpha}}
H^{~'}_{\alpha\acute{\alpha}\gamma\acute{\gamma}} A^{a}_{\gamma\acute{\gamma}},
\eeqa
where
$$
H^{'}_{\alpha\acute{\alpha}\gamma\acute{\gamma}}(k)=
{1 \over 2}(\eta_{\alpha\acute{\gamma}}\eta_{\acute{\alpha}\gamma}
+\eta_{\alpha\acute{\alpha}}\eta_{\gamma\acute{\gamma}})k^2
-{1 \over 2}(\eta_{\alpha\acute{\gamma}}k_{\acute\alpha}k_{\gamma}
+\eta_{\acute\alpha\gamma}k_{\alpha}k_{\acute{\gamma}}
+\eta_{\alpha\acute\alpha}k_{\gamma}k_{\acute{\gamma}}
+\eta_{\gamma\acute{\gamma}}k_{\alpha}k_{\acute\alpha}
-2\eta_{\alpha\gamma}k_{\acute\alpha}k_{\acute{\gamma}}).
$$
It is again invariant with respect to the gauge
transformation $\delta A^{a}_{\mu\lambda} =\partial_{\mu} \xi^{a}_{\lambda}$,
but it is not invariant with respect to the gauge transformations
$\delta A^{a}_{\mu \lambda} =\partial_{\lambda} \eta^{a}_{\mu}$ as one can
see from analogous relations
\be\label{currentdivergenceprime}
k_{\alpha}H^{'}_{\alpha\acute{\alpha}\gamma\acute{\gamma}}(k)=0,~~~
k_{\acute{\alpha}}H^{'}_{\alpha\acute{\alpha}\gamma\acute{\gamma}}(k)=
(k^2 \eta_{\alpha\gamma} -k_{\alpha}k_{\gamma})k_{\acute{\gamma}} \neq 0 .
\ee
As it is obvious from (\ref{currentdivergence}) and
(\ref{currentdivergenceprime}), the total Lagrangian
${{\cal L}}^{free}_2 + {{\cal L}}^{' free}_2$ now  poses new enhanced
invariance with respect to the larger, eight parameter, gauge transformations
\be\label{largegaugetransformation}
\delta A^{a}_{\mu \lambda} =\partial_{\mu} \xi^{a}_{\lambda}+
\partial_{\lambda} \eta^{a}_{\mu} +...,
\ee
where $\xi^{a}_{\lambda}$ and $\eta^{a}_{\mu}$ are eight arbitrary functions, because
\be\label{zeroderivatives}
k_{\alpha}(H_{\alpha\acute{\alpha}\gamma\acute{\gamma}}+
H^{'}_{\alpha\acute{\alpha}\gamma\acute{\gamma}})=0,~~~
k_{\acute{\alpha}}(H_{\alpha\acute{\alpha}\gamma\acute{\gamma}}+
H^{'}_{\alpha\acute{\alpha}\gamma\acute{\gamma}})=0 .
\ee
Thus our free part of the Lagrangian is
\beqa\label{totalfreelagrangian}
{{\cal L}}^{tot~free}_{2} =&-&{1 \over 2}\partial_{\mu}
A^{a}_{\nu \lambda}\partial_{\mu} A^{a}_{\nu \lambda}
+{1 \over 2}\partial_{\mu} A^{a}_{\nu \lambda}\partial_{\nu} A^{a}_{\mu \lambda}+
\nn\\
&+&{1 \over 4} \partial_{\mu} A^{a}_{\nu \lambda} \partial_{\mu } A^{a}_{\lambda\nu}
-{1 \over 4} \partial_{\mu} A^{a}_{\nu \lambda} \partial_{\lambda} A^{a}_{\mu \nu}
-{1 \over 4}\partial_{\nu} A^{a}_{\mu \lambda} \partial_{\mu} A^{a}_{\lambda\nu }
+{1 \over 4} \partial_{\nu } A^{a}_{\mu\lambda} \partial_{\lambda} A^{a}_{\mu \nu}
\nn\\
&+&{1 \over 4}\partial_{\mu} A^{a}_{\nu \nu}\partial_{\mu} A^{a}_{\lambda\lambda}
-{1 \over 2}\partial_{\mu} A^{a}_{\nu \nu} \partial_{\lambda} A^{a}_{\mu\lambda}
+{1 \over 4}\partial_{\nu } A^{a}_{\mu\nu}\partial_{\lambda} A^{a}_{\mu\lambda}
\eeqa
or, in equivalent form, it is
\beqa\label{totfreelagrangianalternativeform}
{{\cal L}}^{tot~free}_{2} ={1 \over 2} A^{a}_{\alpha\acute{\alpha}}
\{(\eta_{\alpha\gamma}\eta_{\acute{\alpha}\acute{\gamma}}
-{1\over 2}\eta_{\alpha\acute{\gamma}}\eta_{\acute{\alpha}\gamma}
-{1\over 2}\eta_{\alpha\acute{\alpha}}\eta_{\gamma\acute{\gamma}})
\partial^{2}
-\eta_{\acute{\alpha}\acute{\gamma}} \partial_{\alpha} \partial_{\gamma}
-\eta_{\alpha\gamma} \partial_{\acute{\alpha}} \partial_{\acute{\gamma}}+\nn\\
+{1\over 2}(\eta_{\alpha\acute{\gamma}} \partial_{\acute{\alpha}} \partial_{\gamma}
+\eta_{\acute{\alpha}\gamma} \partial_{\alpha} \partial_{\acute{\gamma}}
+\eta_{\alpha\acute{\alpha}} \partial_{\gamma} \partial_{\acute{\gamma}}
+\eta_{\gamma\acute{\gamma}} \partial_{\alpha} \partial_{\acute{\alpha}})
\}
A^{a}_{\gamma\acute{\gamma}}
\eeqa
and is invariant with respect to the larger gauge transformations
$
\delta A^{a}_{\mu \lambda} =\partial_{\mu} \xi^{a}_{\lambda}+
\partial_{\lambda} \eta^{a}_{\mu},
$
where $\xi^{a}_{\lambda}$ and $\eta^{a}_{\mu}$ are eight arbitrary functions.
In the momentum representation the quadratic form is
\beqa\label{quadraticform}
H^{tot}_{\alpha\acute{\alpha}\gamma\acute{\gamma}}(k)=
(-\eta_{\alpha\gamma}\eta_{\acute{\alpha}\acute{\gamma}}
+{1 \over 2}\eta_{\alpha\acute{\gamma}}\eta_{\acute{\alpha}\gamma}
+{1 \over 2}\eta_{\alpha\acute{\alpha}}\eta_{\gamma\acute{\gamma}})k^2
+\eta_{\alpha\gamma}k_{\acute\alpha}k_{\acute{\gamma}}
+\eta_{\acute\alpha \acute{\gamma}}k_{\alpha}k_{\gamma}\nn\\
-{1 \over 2}(\eta_{\alpha\acute{\gamma}}k_{\acute\alpha}k_{\gamma}
+\eta_{\acute\alpha\gamma}k_{\alpha}k_{\acute{\gamma}}
+\eta_{\alpha\acute\alpha}k_{\gamma}k_{\acute{\gamma}}
+\eta_{\gamma\acute{\gamma}}k_{\alpha}k_{\acute\alpha}).
\eeqa
In summary, we have the following Lagrangian for the
lower-rank tensor gauge fields:
\beqa\label{totalactiontwo}
{{\cal L}}=  {{\cal L}}_1 +  {{\cal L}}_2 +  {{\cal L}}^{'}_2 =
&-&{1\over 4}G^{a}_{\mu\nu}G^{a}_{\mu\nu}\\
&-&{1\over 4}G^{a}_{\mu\nu,\lambda}G^{a}_{\mu\nu,\lambda}
-{1\over 4}G^{a}_{\mu\nu}G^{a}_{\mu\nu,\lambda\lambda}\nn\\
&+&{1\over 4}G^{a}_{\mu\nu,\lambda}G^{a}_{\mu\lambda,\nu}
+{1\over 4}G^{a}_{\mu\nu,\nu}G^{a}_{\mu\lambda,\lambda}
+{1\over 2}G^{a}_{\mu\nu}G^{a}_{\mu\lambda,\nu\lambda}.\nn
\eeqa
Let us consider the equations of motion which follow from this Lagrangian for
the vector gauge field $A^{a}_{\nu}$:
\beqa\label{equationforfirstranktensor}
&&\nabla^{ab}_{\mu}G^{b}_{\mu\nu}
+{1\over 2 }\nabla^{ab}_{\mu} (G^{b}_{\mu\nu,\lambda\lambda}
- G^{b}_{\mu\lambda,\nu\lambda}
- G^{b}_{\lambda\nu,\mu\lambda})
+ g f^{acb} A^{c}_{\mu\lambda} G^{b}_{\mu\nu,\lambda}\\
&-&{1\over 2 }g f^{acb} (A^{c}_{\mu\lambda} G^{b}_{\mu\lambda,\nu}
+A^{c}_{\lambda\mu} G^{b}_{\mu\nu,\lambda}
+A^{c}_{\mu\nu} G^{b}_{\mu\lambda,\lambda}
-A^{c}_{\lambda\lambda} G^{b}_{\mu\nu,\mu}
-A^{c}_{\mu\lambda\lambda} G^{b}_{\mu\nu}
+ A^{c}_{\mu\lambda\nu} G^{b}_{\mu\lambda})\nn
=0
\eeqa\label{equationforsecondranktensor}
and for the second-rank tensor gauge field $A^{a}_{\nu\lambda}$:
\beqa\label{secondrankfieldequations}
&&\nabla^{ab}_{\mu}G^{b}_{\mu\nu,\lambda}
-{1\over 2} (\nabla^{ab}_{\mu}G^{b}_{\mu\lambda,\nu}
+\nabla^{ab}_{\mu}G^{b}_{\lambda\nu,\mu}
+\nabla^{ab}_{\lambda}G^{b}_{\mu\nu,\mu}
+\eta_{\nu\lambda} \nabla^{ab}_{\mu}G^{b}_{\mu\rho,\rho})\nn\\
&+&g f^{acb} A^{c}_{\mu\lambda} G^{b}_{\mu\nu} -
{1\over 2}g f^{acb}(A^{c}_{\mu\nu} G^{b}_{\mu\lambda}
+A^{c}_{\lambda\mu} G^{b}_{\mu\nu}
+A^{c}_{\mu\mu} G^{b}_{\lambda\nu}
+\eta_{\nu\lambda}  A^{c}_{\mu\rho} G^{b}_{\mu\rho})
=0.
\eeqa
The variation of the action with respect to the third-rank gauge field
$A^{a}_{\nu\lambda\rho}$ will give the equations
\be
\eta_{\lambda\rho}\nabla^{ab}_{\mu}G^{b}_{\mu\nu}-{1\over 2}
(\eta_{\nu\rho}\nabla^{ab}_{\mu}G^{b}_{\mu\lambda}  +
\eta_{\nu\lambda}\nabla^{ab}_{\mu}G^{b}_{\mu\rho}) +
{1\over 2} (\nabla^{ab}_{\rho}G^{b}_{\nu\lambda}  +
\nabla^{ab}_{\lambda}G^{b}_{\nu\rho})=0.
\ee
Representing these system of equations in the form
\beqa\label{perturbativeform}
\partial_{\mu} F^{a}_{\mu\nu,\lambda} = j^{a}_{\nu}\nn\\
\partial_{\mu} F^{a}_{\mu\nu,\lambda}
-{1\over 2} (\partial_{\mu} F^{a}_{\mu\lambda,\nu}
+\partial_{\mu} F^{a}_{\lambda\nu,\mu}
+\partial_{\lambda}F^{a}_{\mu\nu,\mu}
+\eta_{\nu\lambda} \partial_{\mu}F^{a}_{\mu\rho,\rho}) = j^{a}_{\nu\lambda}\nn\\
\eta_{\lambda\rho}\partial_{\mu}F^{a}_{\mu\nu}-{1\over 2}
(\eta_{\nu\rho}\partial_{\mu}F^{a}_{\mu\lambda}  +
\eta_{\nu\lambda}\partial_{\mu}F^{a}_{\mu\rho}) +
{1\over 2} (\partial_{\rho}F^{a}_{\nu\lambda}  +
\partial_{\lambda}F^{a}_{\nu\rho})=
j^{a}_{\nu\lambda\rho},
\eeqa
where $F^{a}_{\mu\nu} =  \partial_{\mu} A^{a}_{\nu  } -
\partial_{\nu} A^{a}_{\mu },~
F^{a}_{\mu\nu,\lambda} = \partial_{\mu} A^{a}_{\nu \lambda} -
\partial_{\nu} A^{a}_{\mu \lambda},~
F^{a}_{\mu\nu,\lambda\rho} = \partial_{\mu} A^{a}_{\nu \lambda\rho} -
\partial_{\nu} A^{a}_{\mu \lambda\rho}$ ,
we can find the corresponding conserved currents
\beqa\label{vectorcurrent}
j^{a}_{\nu } = &-&g f^{abc} A^{b}_{\mu } G^{c}_{\mu\nu }
-g f^{abc}\partial_{\mu} (A^{b}_{\mu } A^{c}_{\nu })\nn\\
&-&{1\over 2 }g f^{abc}A^{b}_{\mu} (G^{c}_{\mu\nu,\lambda\lambda}
- G^{c}_{\mu\lambda,\nu\lambda}
- G^{c}_{\lambda\nu,\mu\lambda})
-{1\over 2 }\partial_{\mu} (I^{a}_{\mu\nu,\lambda\lambda}
- I^{a}_{\mu\lambda,\nu\lambda}
- I^{a}_{\lambda\nu,\mu\lambda})\nn\\
&-& g f^{abc} A^{b}_{\mu\lambda} G^{c}_{\mu\nu,\lambda}
+ {1\over 2 }g f^{abc} (A^{b}_{\mu\lambda} G^{c}_{\mu\lambda,\nu}
+A^{b}_{\lambda\mu} G^{c}_{\mu\nu,\lambda}
+A^{b}_{\mu\nu} G^{c}_{\mu\lambda,\lambda}
-A^{b}_{\lambda\lambda} G^{c}_{\mu\nu,\mu}) \nn\\
&-&{1\over 2} g f^{abc} (A^{b}_{\mu\lambda\lambda} G^{c}_{\mu\nu}
- A^{b}_{\mu\lambda\nu} G^{c}_{\mu\lambda}),
\eeqa
where $I^{a}_{\mu\nu,\lambda\rho}=g f^{abc}(~A^{b}_{\mu}~A^{c}_{\nu\lambda\rho} +
 A^{b}_{\mu\lambda}~A^{c}_{\nu\rho}+A^{b}_{\mu\rho}~A^{c}_{\nu\lambda}
 + A^{b}_{\mu\lambda\rho}~A^{c}_{\nu} ~)$ and
\beqa\label{tensorcurrent}
j^{a}_{\nu\lambda}=&-&g f^{abc} A^{b}_{\mu} G^{c}_{\mu\nu,\lambda}
+{1\over 2 }g f^{abc} (A^{b}_{\mu} G^{c}_{\mu\lambda,\nu}
+A^{b}_{\mu} G^{c}_{\lambda\nu,\mu}
+A^{b}_{\lambda} G^{c}_{\mu\nu,\mu}
+\eta_{\nu\lambda}A^{b}_{\mu} G^{c}_{\mu\rho,\rho})\nn\\
&-&g f^{abc} A^{b}_{\mu\lambda} G^{c}_{\mu\nu} +
{1\over 2}g f^{abc}(A^{b}_{\mu\nu} G^{c}_{\mu\lambda}
+A^{b}_{\lambda\mu} G^{c}_{\mu\nu}
+A^{b}_{\mu\mu} G^{c}_{\lambda\nu}
+\eta_{\nu\lambda}  A^{b}_{\mu\rho} G^{c}_{\mu\rho})\nn\\
&-&g f^{abc} \partial_{\mu}
(A^{b}_{\mu} A^{c}_{\nu\lambda} + A^{b}_{\mu\lambda} A^{c}_{\nu}) +
{1\over 2}g f^{abc}
[\partial_{\mu}(A^{b}_{\mu} A^{c}_{\lambda\nu}+A^{b}_{\mu\nu} A^{c}_{\lambda})
+\partial_{\mu}(A^{b}_{\lambda} A^{c}_{\nu\mu}+A^{b}_{\lambda\mu} A^{c}_{\nu})\nn\\
&+&\partial_{\lambda} (A^{b}_{\mu} A^{b}_{\nu\mu}  + A^{b}_{\mu\mu} A^{c}_{\nu})
+\eta_{\nu\lambda}  \partial_{\mu}
(A^{b}_{\mu} A^{b}_{\rho\rho} + A^{b}_{\mu\rho} A^{c}_{\rho})],
\eeqa
\beqa\label{tensorcurrentthierd}
j^{a}_{\nu\lambda\rho}=&-&\eta_{\lambda\rho} ~g f^{abc} A^{b}_{\mu} G^{c}_{\mu\nu}
+{1\over 2 }g f^{abc} (\eta_{\nu\rho} A^{b}_{\mu} G^{c}_{\mu\lambda}
+\eta_{\nu\lambda}A^{b}_{\mu} G^{c}_{\mu\rho}
-A^{b}_{\rho} G^{c}_{\nu\lambda}
-A^{b}_{\lambda} G^{c}_{\nu\rho})\\
&-&\eta_{\lambda\rho} ~g f^{abc} \partial_{\mu}
(A^{b}_{\mu} A^{c}_{\nu}) +
{1\over 2}g f^{abc}
[\partial_{\mu}(\eta_{\nu\lambda} A^{b}_{\mu} A^{c}_{\rho}
+ \eta_{\nu\rho} A^{b}_{\mu} A^{c}_{\lambda})
-\partial_{\lambda} (A^{b}_{\nu} A^{c}_{\rho})
-\partial_{\rho}(A^{b}_{\nu} A^{c}_{\lambda})].\nn
\eeqa
Thus
\beqa
\partial_{\nu} j^{a}_{\nu}&=&0,~~~\nn\\
\partial_{\nu} j^{a}_{\nu\lambda}&=&0,~~~~
\partial_{\lambda} j^{a}_{\nu\lambda}=0,\nn\\
\partial_{\nu} j^{a}_{\nu\lambda\rho}&=&0,~~~~
\partial_{\lambda} j^{a}_{\nu\lambda\rho}=0,~~~~
\partial_{\rho} j^{a}_{\nu\lambda\rho}=0,
\eeqa
because, as we demonstrated above, the partial derivatives of the l.h.s.
of the equations
(\ref{perturbativeform}) are equal to zero (see equations (\ref{zeroderivatives})
and also equations (\ref{divergencestheird})).

At the linearized level, when the gauge coupling constant g is equal to zero,
the equations of motion (\ref{secondrankfieldequations}) for the second-rank tensor
gauge fields will take the form
\beqa\label{mainequation}
\partial^{2}(A^{a}_{\nu\lambda} -{1\over 2}A^{a}_{\lambda\nu})
-\partial_{\nu} \partial_{\mu}  (A^{a}_{\mu\lambda}-
{1\over 2}A^{a}_{\lambda\mu} )&-&
\partial_{\lambda} \partial_{\mu}  (A^{a}_{\nu\mu} - {1\over 2}A^{a}_{\mu\nu} )
+\partial_{\nu} \partial_{\lambda} ( A^{a}_{\mu\mu}-{1\over 2}A^{a}_{\mu\mu})\nn\\
&+&{1\over 2}\eta_{\nu\lambda} ( \partial_{\mu} \partial_{\rho}A^{a}_{\mu\rho}
-  \partial^{2}A^{a}_{\mu\mu})=0
\eeqa
and, as we shall see below, they describe the propagation of massless particles
of spin 2 and spin 0. First of all it is also easy to see that for the symmetric
part of the tensor gauge field
$(A^{a}_{\nu\lambda} + A^{a}_{\lambda\nu})/2$ our equation
reduces to the well known Fierz-Pauli-Schwinger-Chang-Singh-Hagen-Fronsdal equation
\be\label{fierz}
\partial^{2} A_{\nu\lambda}
-\partial_{\nu} \partial_{\mu}  A_{\mu\lambda} -
\partial_{\lambda} \partial_{\mu}  A_{\mu\nu}
+ \partial_{\nu} \partial_{\lambda}  A_{\mu\mu}
+\eta_{\nu\lambda}  (\partial_{\mu} \partial_{\rho}A_{\mu\rho}
- \partial^{2} A_{\mu\mu}) =0,
\ee
which describes the propagation of massless tensor boson with two
physical polarizations, the $s= \pm 2$ helicity states. For the
antisymmetric part $(A^{a}_{\nu\lambda} - A^{a}_{\lambda\nu})/2$
the equation reduces to the form
\be\label{antisymmetric}
\partial^{2} A_{\nu\lambda}
-\partial_{\nu} \partial_{\mu}  A_{\mu\lambda} +
\partial_{\lambda} \partial_{\mu}  A_{\mu\nu}=0
\ee
and describes the propagation of massless scalar boson with
one physical polarization, the $s= 0$ helicity state.

Alternatively, we can  find out propagating degrees of freedom
directly equation (\ref{mainequation}) in a particular gauge.
Indeed, taking the trace of the equation (\ref{mainequation}) we shall get
\be\label{freeequationtrace}
\partial_{\mu} \partial_{\rho}A^{a}_{\mu\rho}
-  \partial^{2}A^{a}_{\rho\rho} =0,
\ee
and the equation (\ref{mainequation}) takes the form
\beqa\label{tracelessequations}
\partial^{2}(A^{a}_{\nu\lambda} -{1\over 2}A^{a}_{\lambda\nu})
-\partial_{\nu} \partial_{\mu}  (A^{a}_{\mu\lambda}-
{1\over 2}A^{a}_{\lambda\mu} )-
\partial_{\lambda} \partial_{\mu}  (A^{a}_{\nu\mu} - {1\over 2}A^{a}_{\mu\nu} )
+{1\over 2}\partial_{\nu} \partial_{\lambda} A^{a}_{\mu\mu}
=0 .
\eeqa
Using the gauge invariance (\ref{largegaugetransformation}) we can impose the
Lorentz invariant supplementary
conditions on the second-rank gauge fields $A_{\mu\lambda}$:
$
\partial_{\mu} A^{a}_{\mu\lambda} =a_{\lambda} ,~~
\partial_{\lambda} A^{a}_{\mu\lambda} =b_{\mu} ,
$
where $a_{\lambda}$ and $b_{\mu}$ are arbitrary functions, or
one can suggest alternative
supplementary conditions in which the quadratic form
(\ref{totalfreelagrangian}), (\ref{totfreelagrangianalternativeform}),
(\ref{quadraticform}) is diagonal:
\be\label{diagonalgauge}
\partial_{\mu} A^{a}_{\mu\lambda} -{1\over 2} \partial_{\lambda} A^{a}_{\mu\mu}=0,~~
\partial_{\lambda} A^{a}_{\mu\lambda} -{1\over 2} \partial_{\mu}
A^{a}_{\lambda\lambda}=0.
\ee
In this gauge the equation (\ref{tracelessequations}) has the  form
\beqa\label{gaugefixedequations}
\partial^{2} A^{a}_{\nu\lambda}  =0
\eeqa
and in the momentum representation
$
A_{\mu\nu}(x) = e_{\mu\nu}(k) e^{ikx}
$
from  equation (\ref{gaugefixedequations}) it follows that $k^2=0$ and
we have {\it massless particles}.

For the symmetric part of the tensor field $A^{a}_{\mu\lambda}$ the supplementary
conditions (\ref{diagonalgauge}) are equivalent to the harmonic gauge
\be\label{harmonic}
\partial_{\mu} (A^{a}_{\mu\lambda} + A^{a}_{\lambda\mu})
 -{1\over 2} \partial_{\lambda}( A^{a}_{\mu\mu}+A^{a}_{\mu\mu})=0,
\ee
and the residual gauge transformations are defined by the gauge parameters
$ \xi^{a}_{\lambda}+\eta^{a}_{\lambda}$ which should satisfy the equation
\beqa\label{residual1}
\partial^{2}(\xi^{a}_{\lambda}+\eta^{a}_{\lambda})=0.
\eeqa
Thus imposing the harmonic gauge (\ref{harmonic}) and
using the residual gauge transformations
(\ref{residual1}) one can see that the number of propagating physical polarizations
which are described by
the symmetric part of the tensor field $A^{a}_{\mu\lambda}$ are given by two
helicity states $s= \pm 2$ multiplied by the dimension of the group G (a=1,...,N).

For the anisymmetric part of the tensor field $A^{a}_{\mu\lambda}$ the supplementary
conditions (\ref{diagonalgauge}) are equivalent to the Lorentz gauge
\be
\partial_{\mu}(A^{a}_{\mu\lambda} - A^{a}_{\lambda\mu})=0
\ee
and together with the equation of motion they describe the propagation of one
physical polarization of helicity $s= 0$ multiplied by the dimension
of the group G (a=1,...,N).

Thus we have seen that the extended gauge symmetry (\ref{largegaugetransformation})
with eight gauge parameters is sufficient to
exclude all negative norm polarizations from the spectrum of the second-rank
{\it nonsymmetric tensor gauge field} $A_{\mu\lambda}$ which describes now
the propagation of three physical modes of helicities $\pm 2$ and $0$.

In the gauge  (\ref{diagonalgauge})  we shall get
$$
H^{fix}_{\alpha\acute{\alpha}\gamma\acute{\gamma}}(k) =
(\eta_{\alpha\gamma}\eta_{\acute{\alpha}\acute{\gamma}}
-{1\over 2} \eta_{\alpha\acute{\gamma}}\eta_{\acute{\alpha}\gamma}
-{1\over 4}\eta_{\alpha\acute{\alpha}}\eta_{\gamma\acute{\gamma}})(-k^2 )
$$
and the propagator $\Delta_{\gamma\acute{\gamma}\lambda\acute{\lambda}}(k)$
defined by the equation
$$
H^{fix}_{\alpha\acute{\alpha}\gamma\acute{\gamma}}(k)
\Delta_{\gamma\acute{\gamma}\lambda\acute{\lambda}}(k) =
\eta_{\alpha\lambda}\eta_{\acute{\alpha}\acute{\lambda}}~~,
$$
will take the form
\be
\Delta_{\gamma\acute{\gamma}\lambda\acute{\lambda}}(k) = -
{4 \eta_{\gamma\lambda}  \eta_{\acute{\gamma}\acute{\lambda}}
+2 \eta_{\gamma\acute{\lambda}}  \eta_{\acute{\gamma}\lambda}
- 3\eta_{\gamma\acute{\gamma}}\eta_{\lambda\acute{\lambda}}
 \over 3(k^2 - i\varepsilon)}~~.
\ee
The corresponding residue can be represented as a sum
\beqa
{4 \eta_{\gamma\lambda}  \eta_{\acute{\gamma}\acute{\lambda}}
+2 \eta_{\gamma\acute{\lambda}}  \eta_{\acute{\gamma}\lambda}
- 3 \eta_{\gamma\acute{\gamma}}\eta_{\lambda\acute{\lambda}}
 \over 3 }
=&+&(\eta_{\gamma\lambda}\eta_{\acute{\gamma}\acute{\lambda}}
+\eta_{\gamma\acute{\lambda}}\eta_{\acute{\gamma}\lambda}
-\eta_{\gamma\acute{\gamma}}\eta_{\lambda\acute{\lambda}})+\nn\\
&+&{1\over 3}(\eta_{\gamma\lambda}\eta_{\acute{\gamma}\acute{\lambda}}
-\eta_{\gamma\acute{\lambda}}\eta_{\acute{\gamma}\lambda}).
\eeqa
The first term describes the $s= \pm 2$ helicity states and is
represented by the symmetric part of the polarization tensor $e_{\mu\lambda}$,
the second term describes $s= 0$ helicity state and is represented
by its antisymmetric part.
Indeed, for the massless case, when $k_{\mu}$ is aligned along the third axis,
$k_{\mu}= (k,0,0,k)$, we have two independent polarizations of the helicity-2
particle:
\beqa
e^{1}_{\mu\lambda}={1\over \sqrt{2}}
\left( \begin{array}{llll}
  0,0,~~0,0\\
  0,1,~~0,0\\
  0,0,-1,0\\
  0,0,~~0,0
\end{array} \right), e^{2}_{\mu\lambda}={1\over \sqrt{2}}
\left( \begin{array}{ll}
  0,0,0,0\\
  0,0,1,0\\
  0,1,0,0\\
  0,0,0,0
\end{array} \right), \nn\\
\eeqa
with the property that
$
e^{1}_{\gamma\acute{\gamma}}e^{1}_{\lambda\acute{\lambda}}  +
e^{2}_{\alpha\acute{\alpha}} e^{2}_{\gamma\acute{\gamma}}\simeq
{1\over 2}(\eta_{\gamma\lambda}\eta_{\acute{\gamma}\acute{\lambda}}
+\eta_{\gamma\acute{\lambda}}\eta_{\acute{\gamma}\lambda}
-\eta_{\gamma\acute{\gamma}}\eta_{\lambda\acute{\lambda}}).
$
The symbol $\simeq$ means that the equation holds up to longitudinal terms.
The polarization tensor which characterizes the third spin-zero {\it axion}
state has the form
\beqa
e^{A}_{\mu\lambda}={1\over \sqrt{2}}
\left( \begin{array}{ll}
  0,~~0,0,0\\
  0,~~0,1,0\\
  0,-1,0,0\\
  0,~~0,0,0
\end{array} \right),
\eeqa
and
$
e^{A}_{\gamma\acute{\gamma}}e^{A}_{\lambda\acute{\lambda}}
\simeq{1\over 2}(\eta_{\gamma\lambda}\eta_{\acute{\gamma}\acute{\lambda}}
-\eta_{\gamma\acute{\lambda}}\eta_{\acute{\gamma}\lambda}).
$
Thus the general second-rank tensor gauge field with 16
components $A_{\mu\lambda}$ describes in this theory three
physical propagating massless polarizations.

\section{\it Enhanced Local Gauge Algebra }

Let us consider now the symmetries of the remaining two terms in the
full Lagrangian ${{\cal L}}=  {{\cal L}}_1 +  {{\cal L}}_2 +  {{\cal L}}^{'}_2 $.
They have the form
$$
-{1\over 4}G^{a}_{\mu\nu}G^{a}_{\mu\nu,\lambda\lambda}
+{1\over 2} G^{a}_{\mu\nu}G^{a}_{\mu\lambda,\nu\lambda}.
$$
The part which is quadratic in fields  has the form
\beqa
{{\cal L}}^{free}_2 ={1 \over 2} A^{a}_{\alpha }\{&+&
(\eta_{\alpha\gamma} \partial^{2} - \partial_{\alpha} \partial_{\gamma} )
\eta_{\gamma^{'}\gamma^{''}} -
\eta_{\alpha\gamma}\partial_{\gamma^{'}} \partial_{\gamma^{''}}\nn\\
&-&{1 \over 2}
( \eta_{\gamma\gamma^{''}}\partial^{2}-\partial_{\gamma} \partial_{\gamma^{''}} )
\eta_{\alpha\gamma^{'}}
-{1 \over 2}
( \eta_{\gamma\gamma^{'}}\partial^{2}-\partial_{\gamma} \partial_{\gamma^{'}} )
\eta_{\alpha\gamma^{''}}\nn\\
&+&{1 \over 2}\eta_{\gamma\gamma^{''}}\partial_{\alpha} \partial_{\gamma^{'}}
+{1 \over 2}\eta_{\gamma\gamma^{'}}\partial_{\alpha} \partial_{\gamma^{''}}~
\}A^{a}_{\gamma\gamma^{'}\gamma^{''}} \nn\\&=&
{1 \over 2} A^{a}_{\alpha }
H_{\alpha \gamma\gamma^{'}\gamma^{''} } A^{a}_{\gamma\gamma^{'}\gamma^{''} } ,
\eeqa
where the quadratic form in the momentum representation is
\beqa
H_{\alpha \gamma\gamma^{'}\gamma^{''} }(k)=
&-&~~(\eta_{\alpha\gamma}k^2  -k_{\alpha}k_{\gamma})\eta_{\gamma^{'}\gamma^{''}}~
+~~\eta_{\alpha\gamma}k_{\gamma^{'}}k_{\gamma^{''}}\nn\\
&+&{1 \over 2}
( \eta_{\gamma\gamma^{''}}k^{2}-k_{\gamma} k_{\gamma^{''}} )
\eta_{\alpha\gamma^{'}}
-{1 \over 2}\eta_{\gamma\gamma^{''}}k_{\alpha} k_{\gamma^{'}}
\nn\\
&+&{1 \over 2}
( \eta_{\gamma\gamma^{'}}k^{2}-k_{\gamma} k_{\gamma^{'}} )
\eta_{\alpha\gamma^{''}}
-{1 \over 2}\eta_{\gamma\gamma^{'}}k_{\alpha} k_{\gamma^{''}}.
\eeqa
It is useful to represent it in three equivalent forms
\beqa
H_{\alpha \gamma\gamma^{'}\gamma^{''} }(k)=
&-&~~H_{\alpha\gamma}\eta_{\gamma^{'}\gamma^{''}}~
+{1 \over 2}
 H_{\gamma\gamma^{''}} \eta_{\alpha\gamma^{'}}+{1 \over 2}
H_{\gamma\gamma^{'}}\eta_{\alpha\gamma^{''}}\nn\\
&+&{1 \over 2} (\eta_{\alpha\gamma }k_{\gamma^{''}} -
\eta_{\gamma\gamma^{''}}k_{\alpha}  )k_{\gamma^{'}} \nn\\
&+&{1 \over 2} ( \eta_{\alpha\gamma }k_{\gamma^{'}} -
\eta_{\gamma\gamma^{'}}k_{\alpha} )k_{\gamma^{''}} \nn\\
=&-&~~H_{\gamma^{'}\gamma^{''}}\eta_{\alpha\gamma}~
+{1 \over 2}
 H_{\alpha\gamma^{'}} \eta_{\gamma\gamma^{''}}+{1 \over 2}
H_{\alpha\gamma^{''}}\eta_{\gamma\gamma^{'}}\nn\\
&+&{1 \over 2} (\eta_{\gamma^{'}\gamma^{''}}k_{\alpha} -
\eta_{\alpha\gamma^{'}}k_{\gamma^{''}}  )k_{\gamma } \nn\\
&+&{1 \over 2} ( \eta_{\gamma^{'}\gamma^{''}}k_{\alpha} -
\eta_{\alpha\gamma^{''}}k_{\gamma^{'}} )k_{\gamma}\nn\\
=
&-&~~H_{\alpha\gamma}\eta_{\gamma^{'}\gamma^{''}}~
+{1 \over 2}
 H_{\alpha\gamma^{'}} \eta_{\gamma\gamma^{''}}+{1 \over 2}
H_{\alpha\gamma^{''}}\eta_{\gamma\gamma^{'}}\nn\\
&+&{1 \over 2} (\eta_{\alpha\gamma }k_{\gamma^{'}} -
\eta_{\alpha\gamma^{'}}k_{\gamma}  )k_{\gamma^{''}} \nn\\
&+&{1 \over 2} ( \eta_{\alpha\gamma }k_{\gamma^{''}} -
\eta_{\alpha\gamma^{''}}k_{\gamma} )k_{\gamma^{'}} .\nn
\eeqa
As one can see all divergences are equal to zero
\be\label{divergencestheird}
k_{\alpha}H_{\alpha \gamma\gamma^{'}\gamma^{''} }(k)=
k_{\gamma}H_{\alpha \gamma\gamma^{'}\gamma^{''} }(k)=
k_{\gamma^{'}}H_{\alpha \gamma\gamma^{'}\gamma^{''} }(k)=
k_{\gamma^{''}}H_{\alpha \gamma\gamma^{'}\gamma^{''} }(k)=0.
\ee
This result means that the quadratic part of the full Lagrangian
${{\cal L}}=  {{\cal L}}_1 +  {{\cal L}}_2 +  {{\cal L}}^{'}_2 $ is
invariant under the following local gauge transformations
\beqa\label{freedoublepolygaugesymmetric}
\tilde{\delta}_{\eta} A^{a}_{\mu} &=&  \partial_{\mu}\eta^a +... \nonumber\\
(II)~~~~~~~\tilde{\delta}_{\eta} A^{a}_{\mu\nu} &=&   \partial_{\nu}
\eta^{a}_{\mu} +...,\\
\tilde{\delta}_{\eta} A^{a}_{\mu\nu\lambda} &=& \partial_{\nu}
\eta^{a}_{\mu\lambda} + \partial_{\lambda}
\eta^{a}_{\mu\nu} +...\nn\\
........&.&......................................,\nn
\eeqa
in addition to the initial local gauge transformations (\ref{polygauge})
\beqa\label{firstlocal}
\delta_{\xi}  A^{a}_{\mu} &=& \partial_{\mu}\xi^{a}+...  \nonumber\\
(I)~~~~~~~\delta_{\xi}  A^{a}_{\mu\nu} &=&
\partial_{\mu}\xi^{a}_{\nu}+...\nonumber\\
\delta_{\xi}  A^{a}_{\mu\nu\lambda} &=& \partial_{\mu}\xi^{a}_{\nu\lambda}+....
\eeqa

It is important to known how the transformation
(\ref{freedoublepolygaugesymmetric}) looks like
when the gauge coupling constant is not equal to zero. The existence of the full
transformation is guaranteed by the conservation of the corresponding currents
(\ref{vectorcurrent}), (\ref{tensorcurrent}) and (\ref{tensorcurrentthierd}).
At the moment we can only guess
the full form of the second local gauge transformation requiring the closure of the
corresponding algebra. The extension we have found has the form \cite{Savvidy:2005fi}:
\beqa\label{doublepolygaugesymmetric}
\tilde{\delta}_{\eta} A^{a}_{\mu} &=& ( \delta^{ab}\partial_{\mu}
+g f^{acb}A^{c}_{\mu})\eta^b ,~~~~~ \nonumber\\
(II)~~~~~~~\tilde{\delta}_{\eta} A^{a}_{\mu\nu} &=&  ( \delta^{ab}\partial_{\nu}
+  g f^{acb}A^{c}_{\nu})\eta^{b}_{\mu} + g f^{acb}A^{c}_{\mu\nu}\eta^{b},\\
\tilde{\delta}_{\eta} A^{a}_{\mu\nu\lambda} &=& ( \delta^{ab}\partial_{\nu}
+g f^{acb} A^{c}_{\nu})\eta^{b}_{\mu\lambda} +( \delta^{ab}\partial_{\lambda}
+g f^{acb} A^{c}_{\lambda})\eta^{b}_{\mu\nu} \nn\\&~&~~~~~~~~~~~~~~~~~~~~~~~~~~~+
g f^{acb}(  A^{c}_{\mu  \nu}\eta^{b}_{\lambda }+
A^{c}_{\mu \lambda }\eta^{b}_{ \nu}+
A^{c}_{\mu\nu\lambda}\eta^{b})\nn\\
........&.&......................................,\nn
\eeqa
and forms a closed algebraic structure. The composition law of the gauge parameters
$\{ \eta,\eta_{\nu},\eta_{\nu\lambda},... \}$ is the same as in (\ref{gaugealgebra}).

\section{{\it Interaction Vertices}}
We are interested now to analyze the interaction properties
of the tensor gauge bosons prescribed by the gauge principle.
The interaction of the Yang-Mills vector bosons with
the charged tensor gauge bosons is described by the Lagrangian
(\ref{totalactiontwo})
$
{{\cal L}}=  {{\cal L}}_1 +  {{\cal L}}_2 +  {{\cal L}}^{'}_2 .
$
Let us first consider three-particle interaction vertices - VTT.
Explicitly three-linear terms of the Lagrangian ${{\cal L}}_2$
have the form:
\beqa\label{cubicterm}
{{\cal L}}^{cubic}_{2}  =
&-&{1 \over 2}g f^{abc}(\partial_{\mu} A^{a}_{\nu\lambda} -
\partial_{\nu} A^{a}_{\mu\lambda})~ (A^{b}_{\mu}A^{c}_{\nu\lambda}+
A^{b}_{\mu\lambda}A^{c}_{\nu})\nonumber\\
&-&{1 \over 4}g f^{abc}(\partial_{\mu} A^{a}_{\nu} -
\partial_{\nu} A^{a}_{\mu})~ 2A^{b}_{\mu\lambda}A^{c}_{\nu\lambda},
\eeqa
and in addition to the standard Yang-Mills VVV three vector boson interaction vertex
$$
{{\cal L}}^{cubic}_{1} = -{1 \over 2} g f^{abc}(\partial_{\mu} A^{a}_{\nu} -
\partial_{\nu} A^{a}_{\mu})
A^{b}_{\mu} A^{c}_{\nu},
$$
which in the momentum representation has the form
\be
{{\cal V}}^{abc}_{\alpha\beta\gamma}(k,p,q) =
-i g f^{abc} [\eta_{\alpha\beta} (p-k)_{\gamma}+ \eta_{\alpha\gamma} (k-q)_{\beta}
 + \eta_{\beta\gamma} (q-p)_{\alpha}] = -i g f^{abc}
 F_{\alpha\beta\gamma}(k,p,q),
\ee
we have a new three-particle interaction vertex of one vector boson and
two tensor gauge bosons - VTT. In momentum space it has the form
\be
{{\cal V}}^{abc}_{\alpha\acute{\alpha}\beta\gamma\acute{\gamma}}(k,p,q) =
-3 i g f^{abc} [\eta_{\alpha\beta} (p-k)_{\gamma}+ \eta_{\alpha\gamma} (k-q)_{\beta}
 + \eta_{\beta\gamma} (q-p)_{\alpha}] \eta_{\acute{\alpha}\acute{\gamma}}.
\ee
Notice that two parts in (\ref{cubicterm}), which came from  different
terms of the Lagrangian ${{\cal L}}_2$, combine into the VVV vertex and the tensor
$3 \eta_{\acute{\alpha}\acute{\gamma}}$. It is convenient to represent the
vertex in the form
\be
{{\cal V}}^{abc}_{\alpha\acute{\alpha}\beta\gamma\acute{\gamma}}(k,p,q) =
-3 i g f^{abc} F_{\alpha\acute{\alpha}\beta\gamma\acute{\gamma}}(k,p,q).
\ee
We have also a three-particle interaction vertex VTT of one vector boson and
two tensor gauge bosons in the second Lagrangian ${{\cal L}}^{'}_2$. Explicitly the
three-linear terms of Lagrangian ${{\cal L}}^{'}_2$ have the form:

\beqa
{{\cal L}}^{'~cubic}_{2} &=&{1\over 2}g f^{abc}(\partial_{\mu} A^{a}_{\nu\lambda} -
\partial_{\nu} A^{a}_{\mu\lambda})~ (A^{b}_{\mu}A^{c}_{\lambda\nu}+
A^{b}_{\mu\nu}A^{c}_{\lambda})\nonumber\\
&+&{1\over 2}g f^{abc}(\partial_{\mu} A^{a}_{\nu\nu} -
\partial_{\nu} A^{a}_{\mu\nu})~ (A^{b}_{\mu}A^{c}_{\lambda\lambda}+
A^{b}_{\mu\lambda}A^{c}_{\lambda})\nonumber\\
&+&{1\over 2}g f^{abc}(\partial_{\mu} A^{a}_{\nu} -
\partial_{\nu} A^{a}_{\mu})~ (A^{b}_{\mu\nu}A^{c}_{\lambda\lambda}+
A^{b}_{\mu\lambda}A^{c}_{\lambda\nu}) ,
\eeqa
so that in the momentum space we have
\beqa
{{\cal V}}^{'abc}_{\alpha\acute{\alpha}\beta\gamma\acute{\gamma}}(k,p,q)
 &=&{3\over 2} i g f^{abc}
F^{'}_{\alpha\acute{\alpha}\beta\gamma\acute{\gamma}}(k,p,q)\nn\\
F^{'}_{\alpha\acute{\alpha}\beta\gamma\acute{\gamma}}(k,p,q) &=&
(p-k)_{\gamma}(\eta_{\alpha\acute{\gamma}}
\eta_{\acute{\alpha}\beta}+
\eta_{\alpha\acute{\alpha}} \eta_{\beta\acute{\gamma}})\nn\\
&+& (k-q)_{\beta}(\eta_{\alpha\acute{\gamma}} \eta_{\acute{\alpha}\gamma}+
\eta_{\alpha\acute{\alpha}} \eta_{\gamma\acute{\gamma}})\nn\\
&+& (q-p)_{\alpha} (\eta_{\acute{\alpha}\gamma} \eta_{\beta\acute{\gamma}}+
\eta_{\acute{\alpha}\beta} \eta_{\gamma\acute{\gamma}})\nn\\
&+&(p-k)_{\acute{\alpha}}\eta_{\alpha\beta} \eta_{\gamma\acute{\gamma}}+
(p-k)_{\acute{\gamma}} \eta_{\alpha\beta} \eta_{\acute{\alpha}\gamma}\nn\\
&+&(k-q)_{\acute{\alpha}} \eta_{\alpha\gamma} \eta_{\beta\acute{\gamma}}+
(k-q)_{\acute{\gamma}}\eta_{\alpha\gamma} \eta_{\acute{\alpha}\beta}\nn\\
&+&(q-p)_{\acute{\alpha}} \eta_{\beta\gamma} \eta_{\alpha\acute{\gamma}}+
(q-p)_{\acute{\gamma}}\eta_{\alpha\acute{\alpha}} \eta_{\beta\gamma}.
\eeqa
Collecting two terms of the three-point vertex VTT together we shall get
\be
{{\cal V}}^{tot~abc}_{\alpha\acute{\alpha}\beta\gamma\acute{\gamma}}(k,p,q) =
{{\cal V}}^{abc}_{\alpha\acute{\alpha}\beta\gamma\acute{\gamma}}(k,p,q) +
{{\cal V}}^{'abc}_{\alpha\acute{\alpha}\beta\gamma\acute{\gamma}}(k,p,q).
\ee

Let us consider now  four-particle interaction terms of the Lagrangian
${{\cal L}}_1 +  {{\cal L}}_2 +  {{\cal L}}^{'}_2$. We have the
standard  four vector boson interaction vertex VVVV
\beqa
{{\cal V}}^{abcd}_{\alpha\beta\gamma\delta}(k,p,q,r) = - g^2 f^{lac}f^{lbd}
(\eta_{\alpha \beta}
\eta_{\gamma\delta} - \eta_{\alpha \delta} \eta_{\beta\gamma})\nonumber\\
-g^2 f^{lad}f^{lbc} (\eta_{\alpha \beta} \eta_{\gamma\delta} -
\eta_{\alpha \gamma}\eta_{\beta\delta} )\nonumber\\
-g^2 f^{lab}f^{lcd} (\eta_{\alpha \gamma} \eta_{\beta\delta} -
\eta_{\alpha \delta}\eta_{\beta\gamma} )
\eeqa
and a new interaction of two vector and two tensor gauge bosons - the VVTT vertex,
\beqa
{{\cal L}}^{quartic}_{2} =
&-&{1 \over 4}g^2 f^{abc}f^{a\acute{b}\acute{c}}
(A^{b}_{\mu} A^{c}_{\nu\lambda} +
A^{b}_{\mu\lambda} A^{c}_{\nu})(
A^{\acute{b}}_{\mu}A^{\acute{c}}_{\nu\lambda} +
~A^{\acute{b}}_{\mu\lambda}A^{\acute{c}}_{\nu})\nonumber\\
&-&{1 \over 2}g^2 f^{abc}f^{a\acute{b}\acute{c}}
A^{b}_{\mu} A^{c}_{\nu}A^{\acute{b}}_{\mu\lambda}A^{\acute{c}}_{\nu\lambda},
\eeqa
which in the momentum space will take the form
\beqa
{{\cal V}}^{abcd}_{\alpha\beta\gamma\acute{\gamma}\delta\acute{\delta}}(k,p,q,r)=
&-& 6g^2 f^{lac}f^{lbd} (\eta_{\alpha \beta}
\eta_{\gamma\delta} - \eta_{\alpha \delta} \eta_{\beta\gamma})
\eta_{\acute{\gamma}\acute{\delta}}\nonumber\\
&-&6g^2 f^{lad}f^{lbc} (\eta_{\alpha \beta} \eta_{\gamma\delta} -
\eta_{\alpha \gamma}\eta_{\beta\delta} )\eta_{\acute{\gamma}\acute{\delta}}\nonumber\\
&-&6g^2 f^{lab}f^{lcd} (\eta_{\alpha \gamma} \eta_{\beta\delta} -
\eta_{\alpha \delta}\eta_{\beta\gamma} )\eta_{\acute{\gamma}\acute{\delta}}.
\eeqa
The second part of the vertex VVTT comes from the Lagrangian ${{\cal L}}^{'}_2$:
\beqa
{{\cal L}}^{' quartic}_2=&+&{1 \over 4}g^2 f^{abc}f^{a\acute{b}\acute{c}}
(A^{b}_{\mu} A^{c}_{\nu\lambda} +
A^{b}_{\mu\lambda} A^{c}_{\nu})
(A^{\acute{b}}_{\mu}A^{\acute{c}}_{\lambda\nu} +
~A^{\acute{b}}_{\mu\nu}A^{\acute{c}}_{\lambda})\nonumber\\
&+&{1 \over 4}g^2 f^{abc}f^{a\acute{b}\acute{c}}
(A^{b}_{\mu} A^{c}_{\nu\nu} + A^{b}_{\mu\nu} A^{c}_{\nu})
(A^{\acute{b}}_{\mu}A^{\acute{c}}_{\lambda\lambda} +
~A^{\acute{b}}_{\mu\lambda}A^{\acute{c}}_{\lambda})\nonumber\\
&+&{1 \over 2}g^2 f^{abc}f^{a\acute{b}\acute{c}}
A^{b}_{\mu} A^{c}_{\nu}
(A^{\acute{b}}_{\mu\nu}A^{\acute{c}}_{\lambda\lambda} +
A^{\acute{b}}_{\mu\lambda}A^{\acute{c}}_{\lambda\nu}),
\eeqa
which in the momentum representation will take the form
\beqa
{{\cal V}}^{'~abcd}_{\alpha\beta\gamma\acute{\gamma}\delta\acute{\delta}}(k,p,q,r)=
3g^2 f^{lac}f^{lbd} [&+&\eta_{\alpha \beta}
(\eta_{\gamma\acute{\delta}}\eta_{\acute{\gamma}\delta} +
\eta_{\gamma\acute{\gamma}} \eta_{\delta\acute{\delta}})\nn\\
&-&\eta_{\beta\gamma}
(\eta_{\alpha \acute{\delta}}\eta_{\acute{\gamma}\delta} +
\eta_{\alpha\acute{\gamma}} \eta_{\delta\acute{\delta}})\nn\\
&-&\eta_{\alpha\delta}
(\eta_{\beta\acute{\gamma}}\eta_{\gamma\acute{\delta}} +
\eta_{\beta\acute{\delta}} \eta_{\gamma\acute{\gamma}})\nn\\
&+&\eta_{\gamma\delta}
(\eta_{\alpha\acute{\delta}}\eta_{\beta\acute{\gamma}} +
\eta_{\alpha\acute{\gamma}} \eta_{\beta\acute{\delta}})]\nn\\
3g^2 f^{lad}f^{lbc} [&+&\eta_{\alpha \beta}
(\eta_{\gamma\acute{\delta}}\eta_{\acute{\gamma}\delta} +
\eta_{\gamma\acute{\gamma}} \eta_{\delta\acute{\delta}})\nn\\
&-&\eta_{\alpha\gamma}
(\eta_{\beta\acute{\delta}}\eta_{\acute{\gamma}\delta} +
\eta_{\beta\acute{\gamma}} \eta_{\delta\acute{\delta}})\nn\\
&-&\eta_{\beta\delta}
(\eta_{\alpha\acute{\gamma}}\eta_{\gamma\acute{\delta}} +
\eta_{\alpha\acute{\delta}} \eta_{\gamma\acute{\gamma}})\nn\\
&+&\eta_{\gamma\delta}
(\eta_{\alpha\acute{\gamma}}\eta_{\beta\acute{\delta}} +
\eta_{\alpha\acute{\delta}} \eta_{\beta\acute{\gamma}})]\nn\\
3g^2 f^{lab}f^{lcd}[&+&\eta_{\alpha \gamma}
(\eta_{\beta\acute{\gamma}}\eta_{\delta\acute{\delta}} +
\eta_{\beta\acute{\delta}} \eta_{\delta\acute{\gamma}})\nn\\
&-&\eta_{\beta\gamma}
(\eta_{\alpha\acute{\gamma}}\eta_{\delta\acute{\delta}} +
\eta_{\alpha\acute{\delta}} \eta_{\delta\acute{\gamma}})\nn\\
&-&\eta_{\alpha\delta}
(\eta_{\beta\acute{\delta}}\eta_{\gamma\acute{\gamma}} +
\eta_{\beta\acute{\gamma}} \eta_{\gamma\acute{\delta}})\nn\\
&+&\eta_{\beta\delta}
(\eta_{\alpha\acute{\delta}}\eta_{\gamma\acute{\gamma}} +
\eta_{\alpha\acute{\gamma}} \eta_{\gamma\acute{\delta}})].
\eeqa
The total vertex is
\be
{{\cal V}}^{tot~abcd}_{\alpha\beta\gamma\delta}(k,p,q,r)=
{{\cal V}}^{abcd}_{\alpha\beta\gamma\delta}(k,p,q,r) +
{{\cal V}}^{'~abcd}_{\alpha\beta\gamma\delta}(k,p,q,r).
\ee

\section{{\it Third-Rank Tensor Gauge Fields}}
The Lagrangian $ {{\cal L}}_1 +  {{\cal L}}_2 +  {{\cal L}}^{'}_2$ contains the
third-rank gauge fields $A^{a}_{\mu\nu\lambda}$, but without the corresponding
kinetic term. In order to make the fields $A^{a}_{\mu\nu\lambda}$ dynamical
we have added the corresponding Lagrangian ${{\cal L}}_3$ presented at
the second line
of the formula (\ref{firstthreeterms}).
But again the Lagrangian
${{\cal L}}_3$ is not the most general invariant
which can be constructed from the corresponding field strength tensors. There are
seven Lorentz invariant quadratic forms which form the second invariant
Lagrangian ${{\cal L}}^{'}_3$ so that at this level the total Lagrangian is a sum
$$
{{\cal L}}= {{\cal L}}_1 +  {{\cal L}}_2 +
{{\cal L}}^{'}_2 +{{\cal L}}_3+ {{\cal L}}^{'}_3 +...
$$

Indeed, the Lagrangian ${{\cal L}}_3$ has the form (\ref{firstthreeterms}):
\beqa
{{\cal L}}_3 =-{1\over 4}G^{a}_{\mu\nu,\lambda\rho}G^{a}_{\mu\nu,\lambda\rho}
-{1\over 8}G^{a}_{\mu\nu ,\lambda\lambda}G^{a}_{\mu\nu ,\rho\rho}
-{1\over 2}G^{a}_{\mu\nu,\lambda}  G^{a}_{\mu\nu ,\lambda \rho\rho}
-{1\over 8}G^{a}_{\mu\nu}  G^{a}_{\mu\nu ,\lambda \lambda\rho\rho}~,
\eeqa
where the field strength tensors (\ref{fieldstrengthgeneral}) are
\beqa\label{spin4fieldstrenghth}
G^{a}_{\mu\nu ,\lambda \rho \sigma} =
\partial_{\mu} A^{a}_{\nu \lambda \rho \sigma} -
\partial_{\nu} A^{a}_{\mu \lambda\rho\sigma} +
g f^{abc}\{~A^{b}_{\mu}~A^{c}_{\nu \lambda \rho\sigma}
+A^{b}_{\mu\lambda}~A^{c}_{\nu\rho \sigma} +
A^{b}_{\mu\rho }~A^{c}_{\nu\lambda\sigma} +
A^{b}_{\mu\sigma}~A^{c}_{\nu\lambda\rho} +\nn\\
+A^{b}_{\mu\lambda\rho}~A^{c}_{\nu \sigma} +
A^{b}_{\mu\lambda\sigma}~A^{c}_{\nu\rho} +
A^{b}_{\mu\rho\sigma}~A^{c}_{\nu \lambda} +
     A^{b}_{\mu\lambda\rho\sigma }~A^{c}_{\nu} ~\}\nonumber
\eeqa
and
\beqa\label{spin4fieldstrenghth4}
G^{a}_{\mu\nu ,\lambda \rho \sigma\delta} =
\partial_{\mu} A^{a}_{\nu \lambda \rho \sigma\delta} -
\partial_{\nu} A^{a}_{\mu \lambda\rho\sigma\delta} &+&
g f^{abc}\{~A^{b}_{\mu}~A^{c}_{\nu \lambda \rho\sigma\delta}
+\sum_{ \lambda \leftrightarrow \rho,\sigma,\delta}
        A^{b}_{\mu\lambda}~A^{c}_{\nu\rho \sigma\delta} + \nn\\
&+&\sum_{\lambda,\rho \leftrightarrow \sigma,\delta}
        A^{b}_{\mu\lambda\rho}~A^{c}_{\nu\sigma\delta} +
\sum_{\lambda,\rho,\sigma\leftrightarrow \delta}
       A^{b}_{\mu\lambda\rho\sigma}~A^{c}_{\nu\delta} +
     A^{b}_{\mu\lambda\rho\sigma\delta }~A^{c}_{\nu} ~\}.\nonumber
\eeqa
The terms in parenthesis are symmetric over $\lambda \rho\sigma$ and
$\lambda \rho \sigma\delta$ respectively. The Lagrangian ${{\cal L}}_3$
is invariant with respect to the extended gauge transformations  (\ref{polygauge})
of the low-rank gauge fields
$ A_{\mu}, A_{\mu\nu}, A_{\mu\nu\lambda}$  and of the fourth-rank gauge field
(\ref{matrixformofgaugetransformation})
\beqa\label{gaugetransform4}
\delta_{\xi}  A_{\mu\nu\lambda\rho} =\partial_{\mu}\xi_{\nu\lambda\rho}
-i g[A_{\mu},\xi_{\nu\lambda\rho}]
-i g [A_{\mu\nu},\xi_{\lambda\rho}]
-i g [A_{\mu\lambda},\xi_{\nu\rho}]
-i g [A_{\mu\rho},\xi_{\nu\lambda}]-\nn\\
-i g  [A_{\mu\nu\lambda},\xi_{\rho}]
-i g  [A_{\mu\nu\rho},\xi_{\lambda}]
-i g  [A_{\mu\lambda\rho},\xi_{\nu}]
-i g [A_{\mu\nu\lambda\rho},\xi]\nonumber
\eeqa
and also of the fifth-rank tensor gauge field (\ref{matrixformofgaugetransformation})
\beqa\label{gaugetransform5}
\delta_{\xi}  A_{\mu\nu\lambda\rho\sigma} &=&\partial_{\mu}\xi_{\nu\lambda\rho\sigma}
-i g[A_{\mu},\xi_{\nu\lambda\rho\sigma}]
-i g \sum_{\nu \leftrightarrow \lambda\rho\sigma}
[A_{\mu\nu},\xi_{\lambda\rho\sigma}]-\nn\\
   &~&-ig\sum_{\nu\lambda \leftrightarrow \rho\sigma}
   [A_{\mu\nu\lambda},\xi_{\rho\sigma}]
      -ig\sum_{\nu\lambda\rho \leftrightarrow \sigma}
      [A_{\mu\nu\lambda\rho},\xi_{\sigma}]
-i g [A_{\mu\nu\lambda\rho},\xi], ~\nonumber
\eeqa
where the gauge parameters $\xi_{\nu\lambda\rho}$ and $\xi_{\nu\lambda\rho\sigma}$
are totally symmetric rank-3 and rank-4 tensors.
The extended gauge transformation of the higher-rank tensor gauge
fields induces the gauge transformation of the fields strengths of the form
(\ref{variationfieldstrengthgeneral})
\beqa\label{spin4fieldstrenghthtransfor}
\delta G^{a}_{\mu\nu,\lambda\rho\sigma} =
g f^{abc} (~G^{b}_{\mu\nu,\lambda\rho\sigma} ~\xi^c  +
 G^{b}_{\mu\nu, \lambda\rho} ~\xi^{c}_{\sigma}+
G^{b}_{\mu\nu, \lambda\sigma} ~\xi^{c}_{\rho}+
G^{b}_{\mu\nu, \rho\sigma} ~\xi^{c}_{\lambda}+~~~~~~~~~~~~~~~~~~~~~~~~~\nn\\
+ G^{b}_{\mu\nu,\lambda } ~\xi^{c}_{\rho\sigma}+
 G^{b}_{\mu\nu,\rho} ~\xi^{c}_{\lambda\sigma}+
 G^{b}_{\mu\nu,\sigma} ~\xi^{c}_{\lambda \rho}+
G^{b}_{\mu\nu } ~\xi^{c}_{\lambda\rho\sigma}~)\nonumber
\eeqa
and
\beqa\label{fieldstrengh5thtransfor}
\delta G^{a}_{\mu\nu,\lambda\rho\sigma\delta} =
g f^{abc} (~G^{b}_{\mu\nu,\lambda\rho\sigma\delta} ~\xi^c
&+& \sum_{ \lambda\rho,\sigma \leftrightarrow \delta}
G^{b}_{\mu\nu, \lambda\rho\sigma} ~\xi^{c}_{\delta}
+ \nn\\
&+& \sum_{ \lambda\rho \leftrightarrow \sigma,\delta}
G^{b}_{\mu\nu, \lambda\rho} ~\xi^{c}_{\sigma\delta}+
         \sum_{ \lambda \leftrightarrow \rho,\sigma,\delta}
         G^{b}_{\mu\nu, \lambda} ~\xi^{c}_{\rho\sigma\delta}~+
         G^{b}_{\mu\nu } ~\xi^{c}_{\lambda\rho\sigma\delta}).\nonumber
\eeqa
Using the above homogeneous transformations for the field strength
tensors one can demonstrate the invariance of the
Lagrangian ${{\cal L}}_3$ with respect to the extended gauge transformations
(\ref{fieldstrenghparticular}),
(\ref{spin4fieldstrenghthtransfor}) and  (\ref{fieldstrengh5thtransfor})
(see reference \cite{Savvidy:2005zm} for details).

Our purpose now is to present a second invariant Lagrangian
which can be constructed in terms of the above field strength tensors.
Let us consider the following seven Lorentz invariant quadratic forms which
can be constructed by the corresponding field strength tensors
\beqa
G^{a}_{\mu\nu,\lambda\rho}G^{a}_{\mu\lambda,\nu\rho},~~~
G^{a}_{\mu\nu,\nu\lambda}G^{a}_{\mu\rho,\rho\lambda},~~~
G^{a}_{\mu\nu,\nu\lambda}G^{a}_{\mu\lambda,\rho\rho},~~~
G^{a}_{\mu\nu,\lambda}G^{a}_{\mu\lambda,\nu\rho\rho},~~~\nn\\
G^{a}_{\mu\nu,\lambda}G^{a}_{\mu\rho,\nu\lambda\rho},~~~
G^{a}_{\mu\nu,\nu}G^{a}_{\mu\lambda,\lambda\rho\rho},~~~
G^{a}_{\mu\nu}G^{a}_{\mu\lambda,\nu\lambda\rho\rho}.~~~~~~~~~~~~~~~~~
\eeqa
Calculating  the variation of each of these terms with respect to
the gauge transformation (\ref{fieldstrenghparticular}),
(\ref{spin4fieldstrenghthtransfor}) and  (\ref{fieldstrengh5thtransfor})
one can get convinced that the particular linear combination
\beqa\label{actionthreeprime}
{{\cal L}}^{'}_3 &=&  {1\over 4}
G^{a}_{\mu\nu,\lambda\rho}G^{a}_{\mu\lambda,\nu\rho}+
{1\over 4} G^{a}_{\mu\nu,\nu\lambda}G^{a}_{\mu\rho,\rho\lambda}+
{1\over 4}G^{a}_{\mu\nu,\nu\lambda}G^{a}_{\mu\lambda,\rho\rho}\nn\\
&+&{1\over 4}G^{a}_{\mu\nu,\lambda}G^{a}_{\mu\lambda,\nu\rho\rho}
+{1\over 2}G^{a}_{\mu\nu,\lambda}G^{a}_{\mu\rho,\nu\lambda\rho}
+{1\over 4}G^{a}_{\mu\nu,\nu}G^{a}_{\mu\lambda,\lambda\rho\rho}
+{1\over 4}G^{a}_{\mu\nu}G^{a}_{\mu\lambda,\nu\lambda\rho\rho}
\eeqa
forms an invariant Lagrangian.  Indeed, the variation of the first term is
$$
\delta_{\xi} G^{a}_{\mu\nu,\lambda\rho}G^{a}_{\mu\lambda,\nu\rho}=
2g f^{abc}G^{a}_{\mu\nu,\lambda\rho}G^{b}_{\mu\lambda,\nu}\xi^{c}_{\rho}+
2g f^{abc}G^{a}_{\mu\nu,\lambda\rho}G^{b}_{\mu\lambda,\rho}\xi^{c}_{\nu}+
2g f^{abc}G^{a}_{\mu\nu,\lambda\rho}G^{b}_{\mu\lambda}\xi^{c}_{\nu\rho} ,
$$
of the second term is
$$
\delta_{\xi} G^{a}_{\mu\nu,\nu\lambda}G^{a}_{\mu\rho,\rho\lambda}=
2g f^{abc}G^{a}_{\mu\nu,\nu\lambda}G^{b}_{\mu\rho,\rho}\xi^{c}_{\lambda}+
2g f^{abc}G^{a}_{\mu\nu,\nu\lambda}G^{b}_{\mu\rho,\lambda}\xi^{c}_{\rho}+
2g f^{abc}G^{a}_{\mu\nu,\nu\lambda}G^{b}_{\mu\rho}\xi^{c}_{\rho\lambda} ,
$$
of the third term is
\beqa
\delta_{\xi} G^{a}_{\mu\nu,\nu\lambda}G^{a}_{\mu\lambda,\rho\rho}=
2g f^{abc}G^{a}_{\mu\nu,\nu\lambda}G^{b}_{\mu\lambda,\rho}\xi^{c}_{\rho}+
g f^{abc}G^{a}_{\mu\nu,\nu\lambda}G^{b}_{\mu\lambda}\xi^{c}_{\rho\rho}+
g f^{abc}G^{a}_{\mu\lambda,\rho\rho}G^{b}_{\mu\nu,\nu}\xi^{c}_{\lambda}+\nn\\
+g f^{abc}G^{a}_{\mu\lambda,\rho\rho}G^{b}_{\mu\nu,\lambda}\xi^{c}_{\nu}+
g f^{abc}G^{a}_{\mu\lambda,\rho\rho}G^{b}_{\mu\nu}\xi^{c}_{\nu\lambda}\nn ,
\eeqa
of the forth term is
\beqa
\delta_{\xi} G^{a}_{\mu\nu,\lambda}G^{a}_{\mu\lambda,\nu\rho\rho}=
g f^{abc}G^{a}_{\mu\lambda,\nu\rho\rho}G^{b}_{\mu\nu}\xi^{c}_{\lambda}+
2g f^{abc}G^{a}_{\mu\lambda,\nu\rho}G^{b}_{\mu\nu,\lambda}\xi^{c}_{\rho}+
g f^{abc}G^{a}_{\mu\lambda,\rho\rho}G^{b}_{\mu\nu,\lambda}\xi^{c}_{\nu}+\nn\\
+g f^{abc}G^{a}_{\mu\nu,\lambda}G^{b}_{\mu\lambda,\nu}\xi^{c}_{\rho\rho}+
2g f^{abc}G^{a}_{\mu\nu,\lambda}G^{b}_{\mu\lambda,\rho}\xi^{c}_{\nu\rho}+
g f^{abc}G^{a}_{\mu\nu,\lambda}G^{b}_{\mu\lambda}\xi^{c}_{\nu\rho\rho}\nn ,
\eeqa
of the fifth term is
\beqa
\delta_{\xi} G^{a}_{\mu\nu,\lambda}G^{a}_{\mu\rho,\nu\lambda\rho}=\nn\\
g f^{abc}G^{a}_{\mu\rho,\nu\lambda\rho}G^{b}_{\mu\nu}\xi^{c}_{\lambda}+
g f^{abc}G^{b}_{\mu\rho,\nu\lambda}G^{a}_{\mu\nu,\lambda}\xi^{c}_{\rho}+
g f^{abc}G^{b}_{\mu\rho,\nu\rho}G^{a}_{\mu\nu,\lambda}\xi^{c}_{\lambda}+
g f^{abc}G^{b}_{\mu\rho,\lambda\rho}G^{a}_{\mu\nu,\lambda}\xi^{c}_{\nu}+\nn\\
+g f^{abc}G^{a}_{\mu\nu,\lambda}G^{b}_{\mu\rho,\nu}\xi^{c}_{\lambda\rho}+
g f^{abc}G^{a}_{\mu\nu,\lambda}G^{b}_{\mu\rho,\lambda}\xi^{c}_{\nu\rho}+
g f^{abc}G^{a}_{\mu\nu,\lambda}G^{b}_{\mu\rho,\rho}\xi^{c}_{\nu\lambda}+
g f^{abc}G^{a}_{\mu\nu,\lambda}G^{b}_{\mu\rho}\xi^{c}_{\nu\lambda\rho}\nn ,
\eeqa
of the sixth term is
\beqa
\delta_{\xi} G^{a}_{\mu\nu,\nu}G^{a}_{\mu\lambda,\lambda\rho\rho}=\nn\\
g f^{abc}G^{a}_{\mu\lambda,\lambda\rho\rho}G^{b}_{\mu\nu}\xi^{c}_{\nu}+
2g f^{abc}G^{b}_{\mu\lambda,\lambda\rho}G^{a}_{\mu\nu,\nu}\xi^{c}_{\rho}+
g f^{abc}G^{b}_{\mu\lambda,\rho\rho}G^{a}_{\mu\nu,\nu}\xi^{c}_{\lambda}+
g f^{abc}G^{b}_{\mu\lambda,\lambda}G^{a}_{\mu\nu,\nu}\xi^{c}_{\rho\rho}+\nn\\
+2g f^{abc}G^{b}_{\mu\lambda,\rho}G^{a}_{\mu\nu,\nu}\xi^{c}_{\lambda\rho}+
g f^{abc}G^{a}_{\mu\nu,\nu}G^{b}_{\mu\lambda}\xi^{c}_{\lambda\rho\rho}\nn
\eeqa
and finally of the seventh term is
\beqa
\delta_{\xi} G^{a}_{\mu,\nu}G^{a}_{\mu\lambda,\nu\lambda\rho\rho}=\nn\\
2g f^{abc}G^{a}_{\mu\nu}G^{b}_{\mu\lambda,\nu\lambda\rho}\xi^{c}_{\rho}+
g f^{abc}G^{a}_{\mu\nu}G^{b}_{\mu\lambda,\nu\rho\rho}\xi^{c}_{\lambda}+
g f^{abc}G^{a}_{\mu\nu}G^{b}_{\mu\lambda,\lambda\rho\rho}\xi^{c}_{\nu}+
g f^{abc}G^{a}_{\mu\nu}G^{b}_{\mu\lambda,\nu\lambda}\xi^{c}_{\rho\rho}+\nn\\
2g f^{abc}G^{a}_{\mu\nu}G^{b}_{\mu\lambda,\nu\rho}\xi^{c}_{\lambda\rho}+
2g f^{abc}G^{a}_{\mu\nu}G^{b}_{\mu\lambda,\lambda\rho}\xi^{c}_{\nu\rho}+
g f^{abc}G^{a}_{\mu\nu}G^{b}_{\mu\lambda,\rho\rho}\xi^{c}_{\nu\lambda}+\nn\\
g f^{abc}G^{a}_{\mu\nu}G^{b}_{\mu\lambda,\nu}\xi^{c}_{\lambda\rho\rho}+
g f^{abc}G^{a}_{\mu\nu}G^{b}_{\mu\lambda,\lambda}\xi^{c}_{\nu\rho\rho}
+2g f^{abc}G^{a}_{\mu\nu}G^{b}_{\mu\lambda,\rho}\xi^{c}_{\nu\lambda\rho}+
g f^{abc}G^{a}_{\mu\nu}G^{b}_{\mu\lambda}\xi^{c}_{\nu\lambda\rho\rho}.\nn
\eeqa
Some of the terms here are equal to zero, like:
$g f^{abc}G^{a}_{\mu\nu,\lambda}G^{b}_{\mu\rho,\lambda}\xi^{c}_{\nu\rho}$,
$g f^{abc}G^{a}_{\mu\lambda,\lambda}G^{b}_{\mu\nu,\nu}\xi^{c}_{\rho\rho}$
and $g f^{abc}G^{a}_{\mu\nu}G^{b}_{\mu\lambda}\xi^{c}_{\nu\lambda\rho\rho}$.
Amazingly all nonzero terms cancel each other.

In summary, we have the following Lagrangian for the third-rank gauge field
$A^{a}_{\mu\nu\lambda}$:
\beqa\label{actionthreeprime}
{{\cal L}}_3 + c{{\cal L}}^{'}_3
=&-&{1\over 4}G^{a}_{\mu\nu,\lambda\rho}G^{a}_{\mu\nu,\lambda\rho}
-{1\over 8}G^{a}_{\mu\nu ,\lambda\lambda}G^{a}_{\mu\nu ,\rho\rho}
-{1\over 2}G^{a}_{\mu\nu,\lambda}  G^{a}_{\mu\nu ,\lambda \rho\rho}
-{1\over 8}G^{a}_{\mu\nu}  G^{a}_{\mu\nu ,\lambda \lambda\rho\rho}+ \nn\\
&+&{c\over 4}
G^{a}_{\mu\nu,\lambda\rho}G^{a}_{\mu\lambda,\nu\rho}+
{c\over 4} G^{a}_{\mu\nu,\nu\lambda}G^{a}_{\mu\rho,\rho\lambda}+
{c\over 4}G^{a}_{\mu\nu,\nu\lambda}G^{a}_{\mu\lambda,\rho\rho}+\\
&+&{c\over 4}G^{a}_{\mu\nu,\lambda}G^{a}_{\mu\lambda,\nu\rho\rho}
+{c\over 2}G^{a}_{\mu\nu,\lambda}G^{a}_{\mu\rho,\nu\lambda\rho}
+{c\over 4}G^{a}_{\mu\nu,\nu}G^{a}_{\mu\lambda,\lambda\rho\rho}
+{c\over 4}G^{a}_{\mu\nu}G^{a}_{\mu\lambda,\nu\lambda\rho\rho},\nn
\eeqa
where c is an arbitrary constant. As one can now convinced this Lagrangian
coincide with the Lagrangian (\ref{secondfulllagrangian}) when s=2.

In summary at every "level" s we have  two invariant quadratic forms,
they represent a general Lagrangian at level s. The total Lagrangian is a
linear sum of the two Lagrangians $ {{\cal L}}_s + c~ {{\cal L}}^{'}_s $
which are given by formulas (\ref{fulllagrangian1}) and (\ref{secondfulllagrangian}).

\section{{\it Conclusion}}
The transformations considered in the previous sections enlarge
the original algebra of Abelian local gauge transformations
found in \cite{Savvidy:2003fx} (expression (64) in \cite{Savvidy:2003fx}) to
a non-Abelian case
and unify into one multiplet particles with arbitrary spins
and with linearly growing multiplicity.
As we have seen, this leads to a natural generalization of the Yang-Mills theory.
The extended non-Abelian gauge transformations
defined for the tensor gauge fields led to the construction of the
appropriate field strength tensors and of the invariant Lagrangians.
The proposed extension may lead  to a natural inclusion of the standard
theory of fundamental forces into a larger theory in which standard
particles (vector gauge bosons, leptons and quarks) represent a
low-spin subgroup of an enlarged family of particles with higher spins.

As an example of an extended gauge field theory with infinite many gauge fields,
this theory {\it can be associated} with
the field theory of the tensionless strings, because in
our generalization of the non-Abelian Yang-Mills theory we essentially
used the symmetry group which appears as symmetry of the ground state
wave function of the tensionless strings
\cite{Savvidy:2003fx,Savvidy:dv,Savvidy:2005fe}.
Nevertheless I do not know how to
derive it directly from tensionless strings, therefore one can not claim
that they are
indeed identical. The main reason is that the above construction, which
is purely
field-theoretical, has a great advantage of being well defined on and off
the mass-shell,
while the string-theoretical constructions have not been yet developed to
the same level, because the corresponding vertex operators are well defined
only on the mass-shell \cite{Savvidy:2005fe}.
The tensor gauge field theory could probably be
a genuine tensionless string field theory because of the common  symmetry
group, and it would be useful to
understand, whether the string theory can fully reproduce this result.
Discussion of the tensionless strings and related questions can also be
found in \cite{Edgren:2005gq,Turok:2004gb,
Engquist:2005yt,Mourad:2005rt,
Bekaert:2005vh,Brink:2005wh,Gamboa:2004cv,Bonelli:2004ve,Bakas:2004jq,
Bredthauer:2004kv,Gamboa:2003fy,Chagas-Filho:2003wv,Bianchi:2003wx,
Gabrielli:1990ay,Gabrielli:1999xt,Baez:2002jn,Ivanov:1976pg,Ivanov:1979ny,
Curtright:1987zc,Castro:2004hi,Bengtsson:2004cd}.

I would like to thank Ludwig Faddeev for stimulating discussions and his
suggestion to consider the proposed extension of the gauge group as an
example of extended current algebra in analogy with the Kac-Moody current algebra.
My thanks are also to Ioannis Bakas and Thordur Jonsson for helpful
discussions.
This work was partially supported by the EEC Grant no. MRTN-CT-2004-005616.

\vfill
\end{document}